\documentclass{iopart}
\usepackage{graphicx}% eps files
\usepackage{iopams} 
\newlength{\twocolumnwidth}

\newlength{\auxlv}

\newlength{\DL}

\newlength{\pwd} 
\begin{document} 
%******************************************* 
\title[Quantum electrodynamics in response representation]{Causal signal transmission by quantum fields.%
\\ Quantum electrodynamics in response representation.%
} 
%*******************************************
\author{L I Plimak$^1$ and S T Stenholm$^{1,2,3}$} 
\address{$^1$ Institut f\"ur Quantenphysik, Universit\"at Ulm, 
D-89069 Ulm, Germany.} 
\address{$^2$ Physics Department, Royal Institute of Technology, KTH, Stockholm, Sweden.} 
\address{$^3$ Laboratory of Computational Engineering, HUT, Espoo, Finland.} 
%*******************************************
%\author{L.\ I.\ Plimak} 
%\affiliation{\ULM.} 
%*******************************************
%\author{S.\ Stenholm} 
%\affiliation{\ULM.} 
%\affiliation{\StigAffilSw.} 
%\affiliation{\StigAffilFin.} 
%*******************************************
\date{\today} 
%*******************************************
\begin{abstract} 
%*******************************************
Using electromagnetic interaction as an example, response transformations [L.P.\ and S.S., Ann.\ Phys.\ {\bf 323}, 1963, 1989 (2008), {\bf 324}, 600 (2009)] are applied to the standard perturbative approach of quantum field theory. This approach is rewritten in the form where the place of field propagators is taken by the retarded Green function of the field. Unlike in conventional quantum-field-theoretical techniques, the concept of space-time propagation of quantized field is built into our techniques. 
\\
{\em This manuscript is an early version of what later became Refs.\ \cite{DirResp,Maxwell}. Some of the discussions were also included in \cite{RelCaus,RelCausMadrid,ClBehD}. The ``narrow-band'' case remains unpublished.\/}
%******************************************* 
\end{abstract}
%******************************************* 
\pacs{XXZ}
%*******************************************
\maketitle 
%*******************************************
%\tableofcontents 
%**************************************************

%{\em {\bf 
\section*{Note:} This manuscript is an early version of what later became Refs.\ \cite{DirResp,Maxwell}. Some of the discussions were also included in \cite{RelCaus,RelCausMadrid,ClBehD}. The ``narrow-band'' case remains unpublished. 
%\/}
%******************************************* 

%********************************************************
\section{Introduction}\label{ch:IntV}
%********************************************************
In this paper we continue our investigation of dynamical response properties of quantum systems. In papers \cite{API,APII,APIII}, we introduced {\em response transformations\/} of quantum kinematics. In paper \cite{WickCaus}, response transformations were extended to the key technical tool of quantum field theory, Wick's theorem \cite{Wick,Hori,Schweber,VasF}. In this paper, we rewrite in response representation standard perturbative techniques of quantum field theory \cite{Schweber,SchwingerC,Perel,Keldysh}. As a practically important example we consider electromagnetic interactions of light and matter. 

Our approach unifies quantum field theory, phase-space techniques and Kubo's linear response theory, the latter generalised to nonlinear and stochastic response. As was explained in the introduction to a previous paper \cite{WickCaus}, this approach is strictly subject to the Schwinger-Perel-Keldysh type of techniques \cite{SchwingerC,Perel,Keldysh} and does not hold (in fact, cannot be even formulated) in the conventional Feynman-Dyson framework \cite{Schweber}. We make extensive use of functional methods of quantum field theory; for an excellent introduction to these we refer the reader to Vasil'ev's textbook \cite{VasF}. Functional (Hori's \cite{Hori}) form of Wick's theorem is discussed, e.g., by Vasil'ev and in our paper \cite{WickCaus}. 

The paper is structured as follows. In section \ref{ch:B}, we take a ``bird's eye view'' at the physical motivation and the key results of this and two forthcomming papers \cite{PFunc,Sudar}. In section \ref{ch:G}, we summarise formal definitions. In section \ref{ch:I}, we consider the {\em broad-band\/}, or {\em nonresonant\/}, case characterised by the real-field-times-real-current form of electromagnetic interaction. We apply response transformations to Dyson's perturbative techniques. Wick's theorem in its response form (the {\em causal Wick theorem\/} \cite{WickCaus}, for short) emerges naturally in transformed perturbative relations. In the broad-band\ case, these relations stay relatively compact, which allows us to present the formal reasoning more or less in detail. The {\em narrow-band\/}, or {\em resonant\/}, case considered in section \ref{ch:EII} is much bulkier. However, the logic changes little, so we only list the key intermediate relations and final results. In section \ref{ch:UP}, we show that the two cases can be seamlessly merged in a single problem. In \ref{ch:Diag}, we construct a closed diagrammatic solution to the problem of ``dressing'' of matter by electromagnetic interaction. This appendix constitutes a formal ``closure'' of the paper, by associating key dynamical relations with a conventional computational method: Wyld-type diagram techniques \cite{Wyld}. In \ref{ch:UF} we demonstrate that the narrow-band\ case\ is indeed a resonance approximation to the broad-band\ case. The former may be recovered by dropping the counter-rotating terms in formulae for the latter. Although as expected, this result is important for overall consistency. 
%********************************************************
\section{Bird's eye view of macroscopic quantum electrodynamics}%
\label{ch:B}
%*******************************************
\begin{figure}[t]
\begin{center}
\includegraphics[height=75pt]{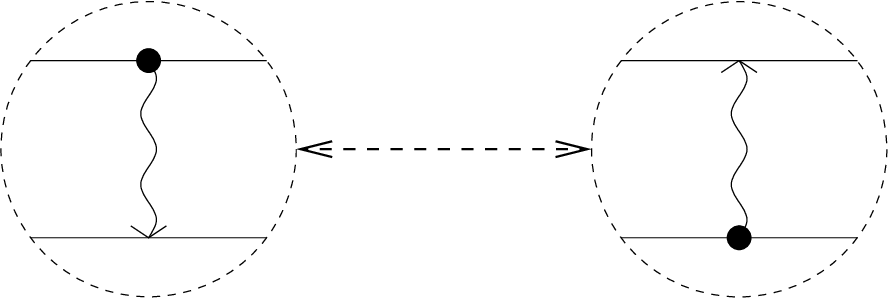}
\end{center}
\caption{Schematics of Fermi's arrangement.}
\label{fig:Fermi}
\end{figure}
%*******************************************
To allow the reader a better insight into our physical motivation, we start from a quick overview of this and two forthcoming papers. A convenient starting point is Fermi's famous arrangement \cite{Fermi,Fano,Arecchi,Milonni,DiffInTr,Nathan} depicted schematically in Fig.\ \ref{fig:Fermi}. Two spatially separated atoms are prepared, one in the ground and the other in the excited state. The atoms are coupled to quantised electromagnetic field. How quickly the state of the second atom would start changing {\em due to the presence of the first atom\/}? The italicized reservation is essential, because the state of the second atom starts changing immediately due to its interaction with the electromagnetic field. Such transients plague any theory with interactions switched on and off --- to which preparation of atoms in their {\em bare\/} ground and excited states is obviously equivalent. One can think of a number of ways how to eliminate this transient. One may look at the difference between transients with the first atom present or absent \cite{DiffInTr}, or improve on the concept of state preparation \cite{Nathan}. However, the most consistent --- and most physical --- way is to adhere to the quantum-field-theoretical viewpoint where all interactions are switched on adiabatically in the remote past \cite{Schweber}. Formally, this means working in the {Heisenberg}\ picture, with all physics expressed in terms of {Heisenberg}\ operators describing field and matter. The price to pay is that state preparation at a finite time is not allowed, and we have to look for more realistic models involving explicit mechanisms of excitation of atoms. 

Rather than formulating a specific model of Fermi's arrangement, we consider a general case of two distinguishable devices coupled to quantised electromagnetic field. 
Such system is described by the generic Hamiltonian, 
%=============================================
{\begin{eqnarray}\eqalign{ 
\hat H(t) = \hbar 
\sum_{\kappa =1}^{N}\omega_{\kappa}\hat a_{\kappa}^{\dag}\hat a_{\kappa} 
+ \hat H_{\mathrm{dev}}(t) + \hat H_{\mathrm{I}}(t) 
. 
}%
\label{eq:86FB} % \nonumber % \Z 
\end{eqnarray}}%
%+++++++++++++++++++++++++++++++++++++++++++++
By definition, we write all Hamiltonian in the interaction picture. The oscillators are represented by the standard bosonic creation and annihilation\ operators, 
%=============================================
{\begin{eqnarray}\eqalign{ 
 \ensuremath{\big[
\hat a_{\kappa},\hat a_{\kappa '}^{\dag}
\big]} = \delta_{k\kappa '} , \quad \kappa ,\kappa'=1,\cdots,N.
}%
\label{eq:87FC} % \nonumber % \Z 
\end{eqnarray}}%
%+++++++++++++++++++++++++++++++++++++++++++++
The nature of the mode index is arbitrary; nothing prevents it from being continuous. The mode frequencies are also arbitrary. The interaction Hamiltonian has the standard field-times-current form, 
%=============================================
{\begin{eqnarray}\eqalign{ 
\hat H_{\mathrm{I}}(t) = -\int dx \hat A(x,t)\hat J(x,t). 
}%
\label{eq:2PW} % \nonumber % \Z 
\end{eqnarray}}%
%+++++++++++++++++++++++++++++++++++++++++++++
The operator of the quantised field $\hat A(x,t)$ is an equally standard linear combination of the creation and annihilation\ operators, 
%=============================================
{\begin{eqnarray}\eqalign{ 
\hat A(x,t) = \sum_{\kappa =1}^{N}\sqrt{\frac{\hbar}{2\omega_{\kappa}}}
 u_{\kappa}(x) \hat a_{\kappa}\mathrm{e}^{-i\omega_{\kappa}t} + \mathrm{H.c.}\, , 
}%
\label{eq:3PX} % \nonumber % \Z 
\end{eqnarray}}%
%+++++++++++++++++++++++++++++++++++++++++++++
where $ u_k(x)$ are complex mode functions. Variable $x$ comprises all field arguments except time. Its meaning, as well as the meaning of the symbol $\int dx$, is problem-specific. The {Heisenberg}\ (initial) quantum state of the system $\hat \rho$ factorises into that of the oscilators and that of the device. The initial state of all oscillators is vacuum, 
%=============================================
{\begin{eqnarray}\eqalign{ 
\hat \rho = \ensuremath{\big| 
0
\big\rangle } 
\ensuremath{\big\langle 
0
\big|} 
\otimes 
\hat \rho_{\mathrm{dev}} 
. 
}%
\label{eq:13ZB} % \nonumber % \Z 
\end{eqnarray}}%
%+++++++++++++++++++++++++++++++++++++++++++++
The device Hamiltonian $\hat H_{\mathrm{dev}}(t)$, the current operator $\hat J(x,t)$ and the state of the device $\hat \rho_{\mathrm{dev}}$ are placeholders for quantum properties of matter. 

In mathematical terms, the Hilbert space of the system is a direct product of the field and matter subspaces. The creation and annihilation\ operators are defined in the former, while $\hat H_{\mathrm{dev}}(t)$, $\hat J(x,t)$ and $\hat \rho_{\mathrm{dev}}$ --- in the latter. The assumption of there being two distinguishable devices is introduced postulating that the matter subspace factorises into two subspaces where the triads $\hat H_{\mathrm{dev}A}(t),\hat J _A(x,t),\hat\rho_{\mathrm{dev}A}$ and $\hat H_{\mathrm{dev}B}(t),\hat J _B(x,t),\hat\rho_{\mathrm{dev}B}$ are defined. These operators characterise macroscopic components of the device, which we refer to as ``device $A$'' and ``device $B$.'' By definition, for operators characterising the composite device we have, 
%=============================================
{\begin{eqnarray} 
\hat H_{\mathrm{dev}}(t) = \hat H_{\mathrm{dev}A}(t)
+\hat H_{\mathrm{dev}B}(t) . 
\label{eq:1FS} % \nonumber % \Z 
\\ 
\hat J(x,t) = \hat J _A(x,t) + \hat J _B(x,t). 
\label{eq:60BC} % \nonumber % \Z 
\\ 
\hat \rho_{\mathrm{dev}} = 
\hat \rho_{\mathrm{dev}A} \otimes 
\hat \rho_{\mathrm{dev}B} . 
\label{eq:62BE} % \nonumber % \Z 
\end{eqnarray}}%
%+++++++++++++++++++++++++++++++++++++++++++++
Apart from most general assumptions of Hermiticity and positivity all operators here are arbitrary. 

%*******************************************
\begin{figure}[t]
\begin{center}
\includegraphics[height=75pt]{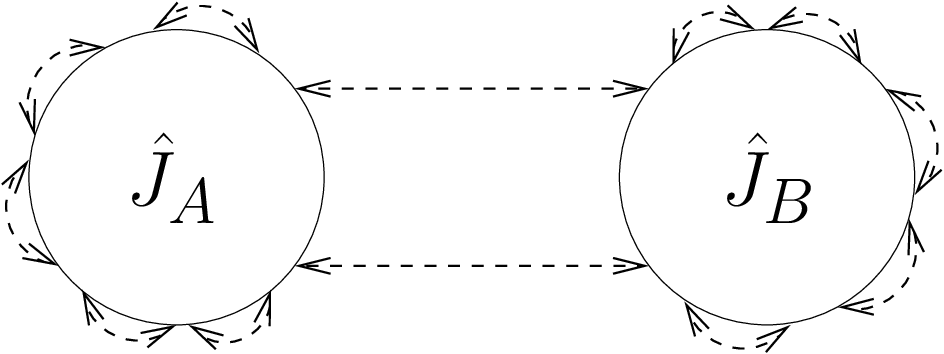}
\end{center}
\caption{Schematics of interactions in the two-device model: conventional quantum approach.}
\label{fig:BirdsEye0}
\end{figure}
%*******************************************
\begin{figure}[t]
\begin{center}
\includegraphics[height=75pt]{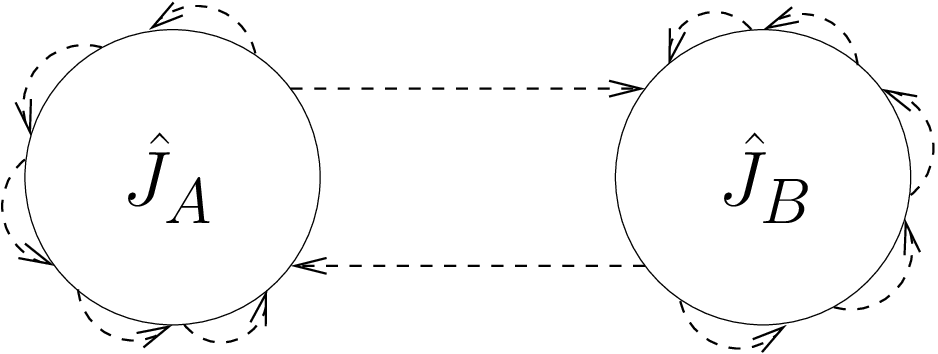}
\end{center}
\caption{Schematics of interactions in the two-device model: field properties are expressed in terms of the retarded Green function.}
\label{fig:BirdsEye3}
\end{figure}
%*******************************************
\begin{figure}[t]
\begin{center}
\includegraphics[height=75pt]{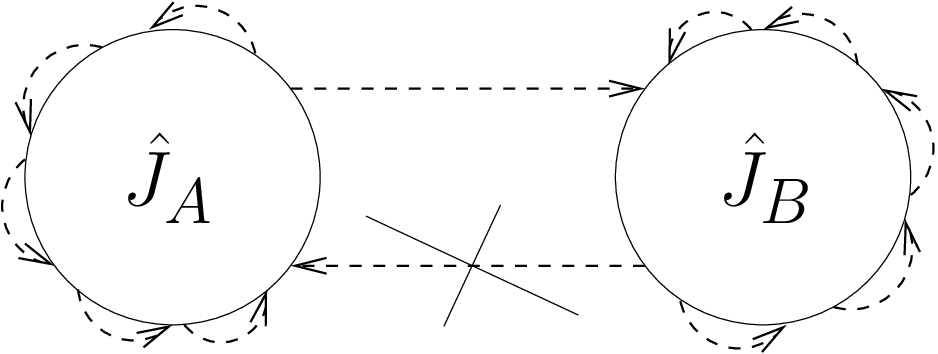}

\vspace{20pt}

\includegraphics[height=75pt]{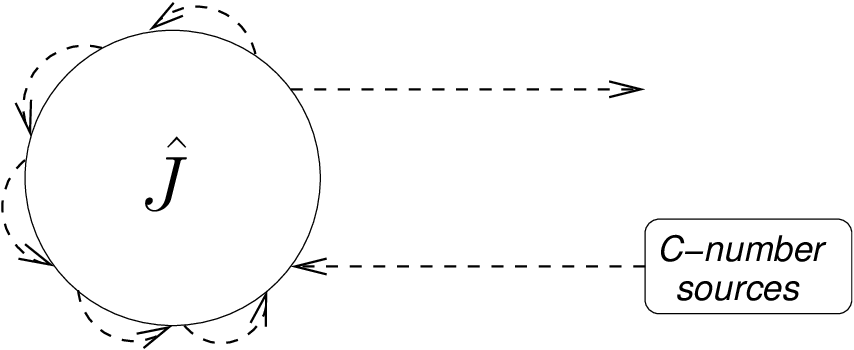}
\end{center}
\caption{Approximations and modifications in the two-device model. Top: neglecting macroscopic back-action leads to generalised photodetection theory. Bottom: full quantum description of a device is restored by adding c-number sources to the model.}
\label{fig:BirdsEye5}
\end{figure}
%*******************************************
Structure of electromagnetic interactions in the outlined model is illustrated in Fig.\ \ref{fig:BirdsEye0}. Assumed distinguishability of the devices leads to separation of the {\em electromagnetic self-action\/} problems for the devices from the problem of their {\em electromagnetic interaction\/}\footnote{%
This statement as well as similar statements below are to be formally verified in this and forthcoming papers \cite{PFunc,Sudar}}. For Fermi's arrangement, this means that one may separate transients from the interaction of atoms via the field (cf.\ also endnote \cite{endTrans}). Furthermore, both in Fig.\ \ref{fig:Fermi} and in Fig.\ \ref{fig:BirdsEye0} the electromagnetic interaction is shown as bidirectional. In Fermi's arrangement, single-directedness is achieved by preparing the atoms in special quantum states (subject to reservations made in endnote \cite{endTrans}). Can one introduce single-directedness in the general arrangement of Fig.\ \ref{fig:BirdsEye0}? 

Formally, directedness of electromagnetic interaction is associated with the linear response function (also known as retarded Green function) of the free electromagnetic field. This quantity emerges if replacing the current operator in (\ref{eq:2PW}) by a c-number source $J_{\mathrm{e}}(x,t)$. This decouples matter from interaction. One can then calculate the linear response of the field \cite{API,Kubo} (see also \cite{SchwingerR}), 
%=============================================
{\begin{eqnarray}\eqalign{ \fl 
\Delta_{\mathrm{R}}(x,x',t-t') = \frac{\delta \ensuremath{\big\langle 
{\hat{\mathcal A}}(x,t)
\big\rangle} }{\delta J_{\mathrm{e}}(x',t')}\settoheight{\auxlv}{$\big|$}%
\raisebox{-0.3\auxlv}{$\big|_{J_{\mathrm{e}}=0}$}
%\\ 
= \frac{i}{\hbar }\theta(t-t')\ensuremath{\big[
\hat A(x,t),\hat A(x',t')
\big]} , 
}%
\label{eq:4PY} % \nonumber % \Z 
\end{eqnarray}}%
%+++++++++++++++++++++++++++++++++++++++++++++
where ${\hat{\mathcal A}}(x,t)$ is the Heisenberg field operator. 
The intermediate expression here is the definition of $\Delta_{\mathrm{R}}$, and the last one is Kubo's formula for it \cite{Kubo}. Kubo's proper expression contains quantum averaging of the commutator, which we omitted because the free-field commutator is a c-number anyway. Kubo's formula may be inverted resulting in the {\em wave quantisation relation\/} \cite{Corresp}, 
%=============================================
{\begin{eqnarray}\eqalign{ \fl 
\ensuremath{\big[
\hat A(x,t),\hat A(x',t')
\big]} %\\ 
= -i\hbar \ensuremath{\big[
\Delta_{\mathrm{R}}(x,x',t-t') - \Delta_{\mathrm{R}}(x',x,t'-t)
\big]} . 
}%
\label{eq:5PZ} % \nonumber % \Z 
\end{eqnarray}}%
%+++++++++++++++++++++++++++++++++++++++++++++
Commutator (\ref{eq:5PZ}) is all one needs to know about the field in order to construct the standard nonstationary perturbation solution to Hamiltonian (\ref{eq:86FB}). Technical means allowing one, firstly, to construct generalised perturbative series without specifying the model of matter, and, secondly, to rewrite it (series) in terms of $\Delta_{\mathrm{R}}$ is rather involved, but conceptually everything remains fairly simple. The resulting structure of electromagnetic interactions is illustrated in Fig.\ \ref{fig:BirdsEye3}. 

The roles the concept of single-directedness plays in the self-action and in the interaction problems are drastically different. In the case of self-action, both ends of $\Delta_{\mathrm{R}}$ are attached, so to speak, to the same electron. There is no way to control microscopic actions and back-actions associated with emitting and reabsorbing the field by matter at a microscopic level. Whereas in case of interaction, the signals travelling from device $A$ to device $B$ may be {\em experimentally distinguished\/} from those travelling from device $B$ to device $A$. This assumption concerns not only physics but also engineering. Physics implies that the situation is macroscopic (often termed mesoscopic, see endnote \cite{endMeso}): devices are distinguishable and separated by macroscopic distances, and all light beams may be controlled. Engineering takes care of such problems as, for instance, preventing light reflected off the detector input window from affecting the laser source. 

Distinguishability of the incoming and outgoing electromagnetic signals for macroscopic electromagnetic devices are put to use in two ways. Dropping the signal travelling from device $B$ to device $A$ yields a generalised photodetection theory (Fig.\ \ref{fig:BirdsEye5}, top). The dropped incoming wave may be restored by introducing c-number external sources into the Hamiltonian (Fig.\ \ref{fig:BirdsEye5}, bottom). 
It is not immediately clear that c-number sources suffice to describe the incoming signal which in arrangement of Fig.\ \ref{fig:BirdsEye3} may be in a quantum state. This statement constitutes a generalisation of Sudarshan's optical equivalence theorem \cite{Sudarshan} to arbitrary interacting systems. 

The said generalisation of Sudarshan's theorem is a particular case of a more general statement. Namely, in the formal structure corresponding to Fig.\ \ref{fig:BirdsEye3}, Planck's constant turns out to be eliminated from all relations between the quantised field and current \cite{Corresp}. Any such relation survives the limit $\hbar \to 0$ unchanged, and must therefore correspond to a particular relation of classical stochastic electrodynamics. The most natural expression of such quantum-classical corrrespondences emerges by mapping quantum relations into phase-space. 

The order in which the aforesaid formal facts will be established is reversed compared to the logic of the above ``pictorial argument.'' The subject of the present paper is a solitary device interacting with c-number sources, as illustrated in Fig.\ \ref{fig:BirdsEye5}, bottom. Phase-space mappings and quantum-classical correspondences will be the subject of paper \cite{PFunc}. Interactions of distinguishable devices, generalised photodetection theory and the optical equivalence theorem will be the subject of paper \cite{Sudar}. Apart from interaction (\ref{eq:2PW}), in all three papers we also look at its modification under the rotating wave approximation, and at how these two types of interaction coexist in a single model.

%********************************************************
\section{Generic model of a macroscopic electromagnetic device}%
\label{ch:G}
%********************************************************
\subsection{RWA or no-RWA?}%
\label{ch:RWA}
%********************************************************
We consider a structural model of electromagnetic interaction where two sets of oscillator modes interact with a ``quantum device.'' The nature of the ``device'' may be arbitrary. For the modes, we assume two different types of coupling: a resonant electrical field-time-dipole coupling for one set, and a nonresonant potential-times-current coupling for the other. Later this will allow us to treat on an equal footing the light (of which the interaction with matter in quantum optics is commonly treated under the resonance, or rotating wave, approximation --- RWA) and the photocurrent (which is a broad-band\ process and in no way is a subject to the RWA). We cannot just assume the RWA because we then lose the low frequency photocurrent modes. Nor are we really keen to keep our analyses without the RWA because this would essentially disconnect us from the quantum-optical paradigm. (To recognise the problem, the reader may try, for example, to reformulate Glauber's photodetection theory \cite{GlauberPhDet,KelleyKleiner,GlauberTN} in terms of real fields rather than analytical signals.) There is no simple notational compromise because the number of dynamical variables in the two cases differ: one real field {\em versus\/} a pair of conjugates. We therefore see no other choice but to develop two versions of the theory in parallel. 

To refer to physics rather than to formal techniques we shall talk about the {\em broad-band\/} (or {\em nonresonant\/}) and the {\em narrow-band\/} (or {\em resonant\/}) cases. Both the response transformations and the causal Wick theorems in the two cases differ. The broad-band\ case\ is a continuation of our papers \cite{API,APII,APIII} (and of paper \cite{Corresp}). The narrow-band\ case\ is an extension of paper \cite{BWO}, see also \cite{Belinicher}. In paper \cite{WickCaus} we referred to these cases as to the real and nonrelativistic ones, respectively. The change of terminology follows the change of motivation: advancing formal techniques in paper \cite{WickCaus} {\em versus\/} understanding macroscopic quantum electrodynamics in this and forthcoming papers. 
%********************************************************
\subsection{The Hamiltonian}%
\label{ch:GH}
%********************************************************
Formally, we assume Hamiltonian (\ref{eq:86FB}) with some amendments. All field oscillators are divided in two groups of $M$ and $N-M$ modes, \mbox{$1\leq M\leq N$}, organised in two electromagnetic-field operators. The {\em narrow-band\/}, or {\em resonant\/}, field is, 
%=============================================
{\begin{eqnarray}\eqalign{ 
\hat E (x,t) = i\sum_{\kappa =1}^{M} 
\sqrt{\frac{\hbar\omega_{\kappa}}{2}} u_{\kappa}(x) \hat a_{\kappa} \mathrm{e}^{-i(\omega_{\kappa}-\omega _0)t}, 
\\ 
\hat E^{\dag} (x,t) = -i\sum_{\kappa =1}^{M} 
\sqrt{\frac{\hbar\omega_{\kappa}}{2}} u^*_{\kappa}(x) \hat a_{\kappa}^{\dag} \mathrm{e}^{i(\omega_{\kappa}-\omega _0)t}, 
}%
\label{eq:4SX} 
\end{eqnarray}}%
%+++++++++++++++++++++++++++++++++++++++++++++
and the {\em broad-band\/}, or {\em nonresonant\/}, field is defined as, 
%=============================================
{\begin{eqnarray}\eqalign{ 
\hat A(x,t) = \sum_{\kappa =M+1}^{N}
\sqrt{\frac{\hbar}{2\omega_{\kappa}}}
 u_{\kappa}(x) \hat a_{\kappa}\mathrm{e}^{-i\omega_{\kappa}t} + \mathrm{H.c.}\, . 
}%
\label{eq:45LB} % \nonumber % \Z 
\end{eqnarray}}%
%+++++++++++++++++++++++++++++++++++++++++++++
In (\ref{eq:4SX}), $\omega _0$ is the characteristic optical carrier frequency. Other quantities in (\ref{eq:4SX}) and (\ref{eq:45LB}) have the same meaning as in section \ref{ch:B}. For simplicity we assume that both fields depend on the same set of arguments. This restriction may be easily lifted. 

For the resonant\ modes, their frequencies $\omega_{\kappa}$ are supposed to occupy a narrow band of frequencies centered at $\omega _0$. That is, $\hat E (x,t)$ and $\hat E^{\dag} (x,t)$ are by definition {\em slow amplitudes\/}. However, this assumption is only important for physics. Formally, it may be disregraded, with the only exception of \ref{ch:UF}. In the rest of the paper, $\omega _0$ is an arbitrary quantity which may safely be set to zero. 

Electromagnetic interaction in (\ref{eq:86FB}) is also divided in two, 
%=============================================
{\begin{eqnarray}\eqalign{ 
\hat H_{\mathrm{I}}(t) = \hat H_{\mathrm{I}} ^{(1)}(t) + \hat H_{\mathrm{I}} ^{(2)}(t) . 
}%
\label{eq:82JT} % \nonumber % \Z 
\end{eqnarray}}%
%+++++++++++++++++++++++++++++++++++++++++++++
The narrow-band\ field interacts with the device 
according to the resonant\ Hamiltonian, 
%=============================================
{\begin{eqnarray}\eqalign{ \fl 
\hat H_{\mathrm{I}} ^{(1)}(t) = 
- \int dx\ensuremath{\big[ 
\hat D (x,t)\hat E ^{\dag}(x,t) + \hat D (x,t)E^*_{\mathrm{e}}(x,t) 
%\\ 
+ D_{\mathrm{e}}(x,t)\hat E ^{\dag}(x,t)
\big]}+\mathrm{H.c.}\, , 
}%
\label{eq:83AR} 
\end{eqnarray}}%
%+++++++++++++++++++++++++++++++++++++++++++++
while the broad-band\ field --- 
according to the nonresonant\ Hamiltonian, 
%=============================================
{\begin{eqnarray}\eqalign{ \fl 
\hat H_{\mathrm{I}}^{(2)}(t) = 
- 
\int dx \ensuremath{\big[
\hat J(x,t)\hat A(x,t)+\hat J(x,t)A_{\mathrm{e}}(x,t)
%\\ 
+ J_{\mathrm{e}}(x,t)\hat A(x,t)
\big]} 
 . 
}%
\label{eq:81AP} % \nonumber % \Z 
\end{eqnarray}}%
%+++++++++++++++++++++++++++++++++++++++++++++
The state of the system is given by equation (\ref{eq:13ZB}). Motivation for introducing c-number sources into the interaction was elucidated in section \ref{ch:B}. 

Comments in the paragraph preceding equation (\ref{eq:1FS}) hold, except there is an additional operator $\hat D (x,t)$ characterising matter. Factorisation of the matter subspace will not be assumed till paper \cite{Sudar}, so that equations (\ref{eq:1FS}), (\ref{eq:60BC}) and (\ref{eq:62BE}) do not apply. 

Fields, currents and dipoles in equations (\ref{eq:81AP}) and (\ref{eq:45LB}) are interaction-picture operators. Their {Heisenberg}\ counterparts will be denoted by calligraphic letters as ${\hat{\mathcal E}}(x,t)$, ${\hat{\mathcal A}}(x,t)$, ${\hat{\mathcal D}}(x,t)$, and ${\hat{\mathcal J}}(x,t)$. 

Hamiltonian (\ref{eq:86FB}) is a placeholder for all conceivable cases of electromagnetic interaction, from a single mode (with $M=N=1$) to relativistic quantum fields (with $M=0$ and $N\to\infty$). From our perspective, this Hamiltonian is a structural model of a quantum-optical experiment involving photodetection. Operators ${\hat{\mathcal E}}(x,t)$ and ${\hat{\mathcal D}}(x,t)$ describe optical interactions, while ${\hat{\mathcal A}}(x,t)$ and ${\hat{\mathcal J}}(x,t)$ express photovoltage and photocurrent. Without specifications, Hamiltonian (\ref{eq:86FB}) is a structural model of photodetector. By breaking the device into distinguishable components, we recover a structural model of the source-detector interaction, etc. 

%********************************************************
\subsection{Operator orderings}%
\label{ch:ADO}
%********************************************************
Here we summarise definitions of operator orderings used in the paper. All definitions imply bosonic operators. Of importance to us are the normal and time orderings. The normal ordering puts $\hat a^{\dag}$'s occuring in equation (\ref{eq:86FB}) to the left of $\hat a$'s, 
%=============================================
{\begin{eqnarray}\eqalign{ 
{\mbox{\rm\boldmath$:$}}
\hat a_{\kappa} \hat a_{\kappa'}^{\dag}
{\mbox{\rm\boldmath$:$}} 
= 
{\mbox{\rm\boldmath$:$}}
\hat a_{\kappa'}^{\dag} \hat a_{\kappa} 
{\mbox{\rm\boldmath$:$}} = \hat a_{\kappa'}^{\dag} \hat a_{\kappa} , 
}%
\label{eq:19WK} % \nonumber % \Z 
\end{eqnarray}}%
%+++++++++++++++++++++++++++++++++++++++++++++
etc. It is extended to the field operators (\ref{eq:4SX}), (\ref{eq:45LB}) by linearity. The time-orderings $T_+$ and $T_-$ put operators in the order of decreasing and increasing time arguments, respectively, %=============================================
{\begin{eqnarray} \fl 
T_+{\hat{\mathcal X}}_1(t_1){\hat{\mathcal X}}_2(t_2)\cdots{\hat{\mathcal X}}_m(t_m) 
\nonumber\\ 
\fl \qquad 
= \sum_{\mathit{perms}} 
\theta(t_1-t_2)
\theta(t_2-t_3)\cdots
\theta(t_{m-1}-t_m) 
%\\ \qquad\times
{\hat{\mathcal X}}_1(t_1){\hat{\mathcal X}}_2(t_2)\cdots{\hat{\mathcal X}}_m(t_m) , 
\label{eq:10TD} % \nonumber % \Z 
\\ \fl
T_-{\hat{\mathcal X}}_1(t_1){\hat{\mathcal X}}_2(t_2)\cdots{\hat{\mathcal X}}_m(t_m) 
\nonumber\\ 
\fl \qquad 
= \sum_{\mathit{perms}} 
\theta(t_1-t_2)
\theta(t_2-t_3)\cdots
\theta(t_{m-1}-t_m) 
{\hat{\mathcal X}}_m(t_m)\cdots{\hat{\mathcal X}}_2(t_2){\hat{\mathcal X}}_1(t_1) . 
\label{eq:11TE} % \nonumber % \Z 
\end{eqnarray}}%
%+++++++++++++++++++++++++++++++++++++++++++++
The sums are over all permutations of times $t_1,\cdots,t_m$. Of use is the property, 
%=============================================
{\begin{eqnarray}\eqalign{ 
{[T_{\pm}{\hat{\mathcal X}}_1(t_1){\hat{\mathcal X}}_2(t_2)\cdots{\hat{\mathcal X}}_m(t_m)]}^{\dag} %\\ 
= T_{\mp}{\hat{\mathcal X}}_1^{\dag}(t_1){\hat{\mathcal X}}_2^{\dag}(t_2)\cdots{\hat{\mathcal X}}_m^{\dag}(t_m). 
}%
\label{eq:94HX} % \nonumber % \Z 
\end{eqnarray}}%
%+++++++++++++++++++++++++++++++++++++++++++++
It expresses the simple fact that Hermitian conjugation inverts the order of factors turning a $T_+$-ordered product into a $T_-$-ordered one, and {\em vice versa\/}. We note in passing that definitions like (\ref{eq:10TD}), (\ref{eq:11TE}) may cause mathematical problems, see the concluding remark in section IIIB of paper \cite{WickCaus}. 

In formulae, the $T_{\pm}$-orderings mostly occur as {\em double time ordered\/} structures $T_-\cdots T_+\cdots$. Rather than visually keeping the operators under the $T_{\pm}$-orderings, one marks the operators with the ${}_{\pm}$ indices and allows them to commute freely: 
%=============================================
{\begin{eqnarray}\eqalign{ 
T_-{\hat{\mathcal X}}_1(t_1)\cdots{\hat{\mathcal X}}_m(t_m) \, 
T_+{\hat{\mathcal Y}}_1(t'_1)\cdots{\hat{\mathcal Y}}_n(t'_n) 
\\ \qquad 
= T_C{\hat{\mathcal X}}_{1-}(t_1)\cdots{\hat{\mathcal X}}_{m-}(t_m) 
{\hat{\mathcal Y}}_{1+}(t'_1)\cdots{\hat{\mathcal Y}}_{n+}(t'_n) 
\\ \qquad = T_C
{\hat{\mathcal Y}}_{1+}(t'_1)\cdots{\hat{\mathcal Y}}_{n+}(t'_n) 
{\hat{\mathcal X}}_{1-}(t_1)\cdots{\hat{\mathcal X}}_{m-}(t_m) , 
}%
\label{eq:10BU} % \nonumber % \Z 
\end{eqnarray}}%
%+++++++++++++++++++++++++++++++++++++++++++++
etc.
The ${}_{\pm}$ indices serve only for ordering purposes and otherwise should be disregarded. 
Using $T_C$ results in a reduction of formulae in bulk which may be truly dramatic (compare, e.g., expressions for response functions in \cite{APII} and in \cite{APIII}). In this paper we mostly use the $T_C$-ordering and revert to the double time ordering only where necessary. 

%********************************************************
\subsection{Quantum Green functions}%
\label{ch:QG}
%********************************************************
Quantized fields $\hat E (x,t)$ and $\hat A(x,t)$ enter all formulae through their {\em Green functions\/}. In either case, we define the retarded Green function and two contractions: the Feynman propagator, and one more kernel which emerges as a propagator in Perel-Keldysh's diagram techniques (and which, strictly speaking, is not a Green function). For the resonant\ field, these quantities read, respectively, 
%=============================================
{\begin{eqnarray} 
 G_{\mathrm{R}}(x,x',t-t') = \frac{i}{\hbar}\theta(t-t')\ensuremath{\big\langle 0\big| 
\ensuremath{\big[
\hat E(x,t),\hat E^{\dag}(x',t')
\big]} 
\big|0\big\rangle} , 
\label{eq:85LJ} % \nonumber % \Z 
\\ 
 G_{\mathrm{F}}(x,x',t-t') = \frac{i}{\hbar}\ensuremath{\big\langle 0\big|
T_+\hat E (x,t)\hat E ^{\dag}(x',t')
\big|0\big\rangle} , 
\label{eq:86LK} % \nonumber % \Z 
\\
 G^{(+)}(x,x',t-t') = \frac{i}{\hbar}\ensuremath{\big\langle 0\big|
\hat E (x,t)\hat E ^{\dag}(x',t')
\big|0\big\rangle} . 
\label{eq:87LL} % \nonumber % \Z 
\end{eqnarray}}%
%+++++++++++++++++++++++++++++++++++++++++++++
For the nonresonant\ field, they are,
%=============================================
{\begin{eqnarray} 
\label{eq:83LF} % \nonumber % \Z 
\Delta_{\mathrm{R}}(x,x',t-t') = \frac{i}{\hbar }\theta(t-t') \ensuremath{\big\langle 0\big| 
\ensuremath{\big[
\hat A(x,t),\hat A(x',t')
\big]} \big|0\big\rangle}, \\ 
\label{eq:84LH} % \nonumber % \Z 
\Delta _{\mathrm{F}}(x,x',t-t') = \frac{i}{\hbar }\ensuremath{\big\langle 0\big|
T_+\hat A(x,t)\hat A(x',t')
\big|0\big\rangle} , \\ 
\Delta^{(+)}(x,x',t-t') = \frac{i}{\hbar }\ensuremath{\big\langle 0\big|
\hat A(x,t)\hat A(x',t')
\big|0\big\rangle} . 
\label{eq:81LD} % \nonumber % \Z 
\end{eqnarray}}%
%+++++++++++++++++++++++++++++++++++++++++++++
$T_+$ in (\ref{eq:86LK}) and (\ref{eq:84LH}) is the standard time-ordering defined by equation (\ref{eq:10TD}). Using (\ref{eq:4SX}), (\ref{eq:45LB}) we find the explicit formulae, 
%=============================================
{\begin{eqnarray}\eqalign{ 
G^{(+)}(x,x',t-t') = i\sum_{\kappa =1}^{M}
\frac{\omega_{\kappa} u_{\kappa}(x) u_{\kappa}^*(x')\mathrm{e}^{-i\omega_{\kappa}(t-t')}}
{2}, \\ 
\Delta^{(+)}(x,x',t-t') = 
i\sum_{\kappa =M+1}^{N}\frac{ u_{\kappa}(x) u_{\kappa}^*(x')\mathrm{e}^{-i\omega_{\kappa}(t-t')}}
{2\omega_{\kappa}} . 
}%
\label{eq:20FA} % \nonumber % \Z 
\end{eqnarray}}%
%+++++++++++++++++++++++++++++++++++++++++++++
Other kernels follow from the relations, 
%=============================================
{\begin{eqnarray} 
\fl \eqalign{
G_{\mathrm{R}}(x,x',t-t') = G_{\mathrm{F}}(x,x',t-t') 
%\np 
= \theta(t-t')G^{(+)}(x,x',t-t') , \\ 
\Delta_{\mathrm{R}}(x,x',t-t')
%\np 
= \theta(t-t')\ensuremath{\big[
\Delta ^{(+)}(x,x',t-t') -\Delta ^{(+)}(x',x,t'-t)
\big]} , \\ 
\Delta _{\mathrm{F}}(x,x',t-t') %\np 
= \theta(t-t')
\Delta ^{(+)}(x,x',t-t') +\theta(t'-t)\Delta ^{(+)}(x',x,t'-t)
. 
}
\label{eq:88LM} % \nonumber % \Z 
\end{eqnarray}}%
%+++++++++++++++++++++++++++++++++++++++++++++

Definitions of the retarded Green functions (\ref{eq:85LJ}) and (\ref{eq:83LF}) are Kubo's formulae for linear response functions \cite{Kubo}. Definition of $\Delta_{\mathrm{R}}$ was discussed in section \ref{ch:B}, cf.\ equation (\ref{eq:4PY}) and comments thereon. $G_{\mathrm{R}}$ is defined in a similar manner, setting \mbox{$
\hat D(x,t)=0
$} in Hamiltonian (\ref{eq:83AR}) and then defining the linear response of the resonant\ field by the formula, 
%=============================================
{\begin{eqnarray}\eqalign{ 
G_{\mathrm{R}}(x,x',t-t') = \frac{\delta\ensuremath{\big\langle 
{\hat{\mathcal E}}(x,t)
\big\rangle} }{\delta D_{\mathrm{e}}(x',t')} 
\settoheight{\auxlv}{$\Big|$}%
\raisebox{-0.3\auxlv}{$\Big|_{D_{\mathrm{e}}=0}$}. 
}%
\label{eq:6QA} % \nonumber % \Z 
\end{eqnarray}}%
%+++++++++++++++++++++++++++++++++++++++++++++
Equation (\ref{eq:85LJ}) is Kubo's formula for $G_{\mathrm{R}}$. For more details on the linear response theory for free bosonic fields see papers \cite{API,SchwingerR}. Commutators in (\ref{eq:85LJ}) and (\ref{eq:83LF}) are c-numbers, so that the averaging may be dropped (in other words, response of a linear system does not depend on its state). In the other four definitions, this averaging matters. 

%********************************************************
\subsection{Response transformation of contractions}%
\label{ch:GR}
%********************************************************
All our results hinge on one observation: the propagators (contractions) may be expressed by the retarded Green functions \cite{API,WickCaus,Corresp,BWO}. In the narrow-band\ case, this connection readily follows from (\ref{eq:88LM}), 
%=============================================
{\begin{eqnarray}\eqalign{ 
G_{\mathrm{F}}(x,x',t-t') = G_{\mathrm{R}}(x,x',t-t'), \\ 
G^{(+)}(x,x',t-t') = G_{\mathrm{R}}(t-t')- G_{\mathrm{R}}^*(x,x',t'-t). 
}%
\label{eq:50UX} % \nonumber % \Z 
\end{eqnarray}}%
%+++++++++++++++++++++++++++++++++++++++++++++
In the broad-band\ case, such connection is more involved. It happens to be an integral transformation \cite{API,WickCaus,BWO}, 
%=============================================
{\begin{eqnarray}\eqalign{ 
\Delta _{\mathrm{F}}(x,x',t-t') 
%\Z \np 
=\Delta_{\mathrm{R}}^{(+)}(x,x',t-t') +\Delta_{\mathrm{R}}^{(+)}(x',x,t'-t), \\ 
\Delta^{(+)}(x,x',t-t') 
%\Z \np 
=\Delta_{\mathrm{R}}^{(+)}(x,x',t-t') -\Delta_{\mathrm{R}}^{(-)}(x',x,t'-t). 
}%
\label{eq:11JQ} % \nonumber % \Z 
\end{eqnarray}}%
%+++++++++++++++++++++++++++++++++++++++++++++
The symbols ${}^{(\pm)}$ denote separation of the frequency-positive and negative\ parts of functions, 
%=============================================
{\begin{eqnarray}\eqalign{ 
 f(t) = f^{(+)}(t) + f^{(-)}(t) , 
\\ 
 f^{(\pm)}(t) = \int_{-\infty}^{+\infty} \frac{d\omega }{2\pi }\mathrm{e}^{-i\omega t}\theta(\pm\omega )
f_{\omega }, 
\\ f_{\omega } = \int_{-\infty}^{+\infty} dt \mathrm{e}^{i\omega t}f(t) . 
}%
\label{eq:4JH} % \nonumber % \Z 
\end{eqnarray}}%
%+++++++++++++++++++++++++++++++++++++++++++++
The frequency-positive and negative\ parts are always defined with respect to native arguments of functions; so,
\mbox{$\Delta_{\mathrm{R}}^{(\pm)}(x',x,t'-t)$} in (\ref{eq:11JQ}) are deciphered as 
\mbox{$\Delta_{\mathrm{R}}^{(\pm)}(x',x,t)\settoheight{\auxlv}{$|$}%
\raisebox{-0.3\auxlv}{$|_{t\to t'-t}$}$}. For more details on this operation see paper \cite{APII}, appendix A. Note that notation for $\Delta ^{(+)}$ and $G^{(+)}$ is consistent with them being purely frequency-positive.

Equations (\ref{eq:11JQ}) may be traced down to the wave quantisation formula (\ref{eq:5PZ}), while (\ref{eq:50UX}) --- to a similar formula for the narrow-band\ field, 
%=============================================
{\begin{eqnarray}\eqalign{ 
\ensuremath{\big[
\hat E(x,t),\hat E^{\dag}(x',t')
\big]} %\\ 
= -i\hbar \ensuremath{\big[
G_{\mathrm{R}}(x,x',t-t') - G^*_{\mathrm{R}}(x',x,t'-t)
\big]} . 
}%
\label{eq:7QB} % \nonumber % \Z 
\end{eqnarray}}%
%+++++++++++++++++++++++++++++++++++++++++++++
Note that $\Delta_{\mathrm{R}}$ is real, while $G_{\mathrm{R}}$ is complex. 
For more detals see papers \cite{API,WickCaus,BWO}. 

%********************************************************
\subsection{Condensed notation}%
\label{ch:DN}
%********************************************************
To keep the bulk of formulae under the lid and make their structure more transparent, we make extensive use of condensed notation, 
%=============================================
{\begin{eqnarray} 
 fg = \int dx dt f(x,t) g(x,t) , 
\label{eq:3VS} % \nonumber % \Z 
\\
 fKg = \int dx dx' dt dt' f(x,t)K(x,x',t-t') 
%\\ \times 
g(x',t') , 
\label{eq:76NU} % \nonumber % \Z 
\\ 
 (Kg)(x,t) = \int dx' dt' K(x,x',t-t')g(x',t') , 
\label{eq:77NV} % \nonumber % \Z 
\\ 
 (fK)(x,t) = \int dx' dt' g(x',t')K(x',x,t'-t) , 
\label{eq:78NW} % \nonumber % \Z 
\end{eqnarray}}%
%+++++++++++++++++++++++++++++++++++++++++++++
where $f(x,t)$ and $g(x,t)$ are c-number or q-number functions, and $K(x,x',t-t')$ is a c-number kernel. The ``products'' $fg$ and $fKg$ denote scalars, while $Kg$ and $fK$ --- functions (fields). 

% temphide

%********************************************************
\section{Oscillators interacting with quantised current 
(the broad-band\ case)}%
\label{ch:I}
%********************************************************
\subsection{The closed-time-loop formalism}%
\label{ch:IC}
%********************************************************
Perturbative formulae for Hamiltonian (\ref{eq:86FB}) grow insufferably bulky. We therefore start from the relatively compact broad-band\ case. Formally, this corresponds to setting \mbox{$M=0$} in equations (\ref{eq:86FB})--(\ref{eq:45LB}). 

To maintain connection with papers \cite{API,APII,APIII,Corresp,BWO}, we solve for the characteristic functional of products of the Heisenberg field and current operators, ${\hat{\mathcal A}}(x,t)$ and ${\hat{\mathcal J}}(x,t)$, ordered in the Schwinger-Perel-Keldysh {\em closed-time-loop\/} style \cite{SchwingerC,Perel,Keldysh}, 
%=============================================
{\begin{eqnarray}\eqalign{ 
\Xi ({\eta}_+,{\eta}_-,{\zeta}_+,{\zeta}_-
| A_{\mathrm{e}} ,J_{\mathrm{e}}) 
\\ \qquad 
= \ensuremath{\big\langle
T_-\exp\ensuremath{\big(
-i{\eta}_-{\hat{\mathcal A}} 
-i{\zeta}_-{\hat{\mathcal J}}
 \big)} 
%\\ \times
\,T_+\exp\ensuremath{\big(
i{\eta}_+{\hat{\mathcal A}} 
+i{\zeta}_+{\hat{\mathcal J}} 
 \big)} 
\big\rangle} 
\\ \qquad 
= \ensuremath{\big\langle
T_C\exp\ensuremath{\big(
i{\eta}_+{\hat{\mathcal A}}_+ 
-i{\eta}_-{\hat{\mathcal A}}_- 
+i{\zeta}_+{\hat{\mathcal J}}_+ 
-i{\zeta}_-{\hat{\mathcal J}}_-
 \big)} 
\big\rangle} 
, 
}%
\label{eq:2XC} % \nonumber % \Z 
\end{eqnarray}}%
%+++++++++++++++++++++++++++++++++++++++++++++
where $\eta_{\pm}(x,t)$ and $\zeta_{\pm}(x,t)$ are four independent complex functions. We use condensed notation (\ref{eq:3VS}). The averaging in (\ref{eq:2XC}) is over the initial (Heisenberg) state of the system, 
%=============================================
{\begin{eqnarray}\eqalign{ 
\ensuremath{\big\langle 
\cdots
\big\rangle} = \mathrm{Tr} \hat \rho (\cdots), 
}%
\label{eq:12ZA} % \nonumber % \Z 
\end{eqnarray}}%
%+++++++++++++++++++++++++++++++++++++++++++++
where $\hat \rho$ is given by equation (\ref{eq:13ZB}) and the ellipsis stands for an arbitrary operator. Definitions of operator orderings were summarised in section \ref{ch:ADO}. 

In this paper, we denote various characteristic functionals of $T_C$-ordered operator averages as $\Xi$, while the same functionals in causal variables will be denoted as $\Phi$. This agrees with notational conventions of paper \cite{BWO}. Notation used in papers \cite{API,APII,APIII} is recovered replacing $\Xi \to\Phi $ and $\Phi \to\Phi_{\mathrm{R}}$. The reason for the change of notation compared to \cite{API,APII,APIII} is obvious: we attach quite a number of indices to the functionals, and an additional index would be a nuisance. 
%********************************************************
\subsection{Dyson's perturbative technique}%
\label{ch:DPA}
%********************************************************
As in papers \cite{Corresp,BWO} we employ Dyson's standard perturbative techniques \cite{Schweber}. We assume that the reader is familiar with the concept of S-matrix and its representation as a T-exponent, 
%=============================================
{\begin{eqnarray}\eqalign{ 
{\hat{\mathcal S}} = T_+\exp\ensuremath{\bigg[
-\frac{i}{\hbar } \int dt \hat H (t)
\bigg]} , 
}%
\label{eq:50ZE} % \nonumber % \Z 
\end{eqnarray}}%
%+++++++++++++++++++++++++++++++++++++++++++++
where $\hat H (t)$ is the interaction Hamiltonian in the interaction picture. As in papers \cite{API,APII,APIII,WickCaus}, omitted integration limits imply the maximal possible area of integration: the whole time axis, the whole space, etc.

The key relation for us is the formula expressing time-ordered products of {Heisenberg}\ operators, ${\hat{\mathcal X}}_1(t_1),\cdots,{\hat{\mathcal X}}_m(t_m)$, as time-ordered products of the same operators in the interaction picture, $\hat{X}_1(t_1),\cdots,\hat{X}_m(t_m)$,
%=============================================
{\begin{eqnarray}\eqalign{ 
T_+ {\hat{\mathcal X}}_1(t_1)\cdots{\hat{\mathcal X}}_m(t_m)
= {\hat{\mathcal S}}^{\dag} T_+{\hat{\mathcal S}}\hat{X}_1(t_1)\cdots\hat{X}_m(t_m). 
}%
\label{eq:93HW} % \nonumber % \Z 
\end{eqnarray}}%
%+++++++++++++++++++++++++++++++++++++++++++++
Detailed derivation of this formula may be found, for instance, in Schweber's textbook \cite{Schweber}. Uncharacteristically for ever-so-thorough Schweber, derivation in \cite{Schweber} disregards the ${\hat{\mathcal S}}^{\dag} $ factor; for the necessary amendments see our paper \cite{APIII}, section 5.2. Taking the adjoint of equation (\ref{eq:93HW}) and using equation (\ref{eq:94HX}) we also obtain, 
%=============================================
{\begin{eqnarray}\eqalign{ 
T_- {\hat{\mathcal X}}_1(t_1)\cdots{\hat{\mathcal X}}_m(t_m)
= \ensuremath{\big[
T_-{\hat{\mathcal S}}^{\dag}\hat{X}_1(t_1)\cdots\hat{X}_m(t_m)
\big]} {\hat{\mathcal S}}. 
}%
\label{eq:95HY} % \nonumber % \Z 
\end{eqnarray}}%
%+++++++++++++++++++++++++++++++++++++++++++++
Square brackets mark the region to which the ordering applies. 

Equations (\ref{eq:50ZE})--(\ref{eq:95HY}) imply that the S-matrix is regarded a functional of the operators $\hat A (x,t)$ and $\hat J (x,t)$. The reader should keep this in mind to avoid confusion and misunderstanding. For details see Vasil'ev \cite{VasF}. 

In the case at hand, the S-matrix and its inverse read, 
%=============================================
{\begin{eqnarray} 
{\hat{\mathcal S}} = T_+\exp\ensuremath{\bigg[
\frac{i}{\hbar }\ensuremath{\big(
\hat A\hat J+ A_{\mathrm{e}}\hat J+ \hat A J_{\mathrm{e}}
 \big)}
\bigg]} , 
\label{eq:6XH} % \nonumber % \Z 
\\
{\hat{\mathcal S}}^{\dag} = T_-\exp\ensuremath{\bigg[
-\frac{i}{\hbar }\ensuremath{\big(
\hat A\hat J+ A_{\mathrm{e}}\hat J+\hat A J_{\mathrm{e}}
 \big)}
\bigg]} ,
\label{eq:56ZM} % \nonumber % \Z 
\end{eqnarray}}%
%+++++++++++++++++++++++++++++++++++++++++++++
where $\hat A (x,t)$ and $\hat J (x,t)$ are the field and current operators in the interaction picture. The former is given by (\ref{eq:45LB}), while the latter is just assumed to be known. We continue using condensed notation (\ref{eq:3VS}). Applying equations (\ref{eq:93HW}), (\ref{eq:95HY}) we obtain, 
%=============================================
{\begin{eqnarray} 
\fl T_+
\exp\ensuremath{\big(
i\eta_+{\hat{\mathcal A}}+i\zeta_+{\hat{\mathcal J}}
 \big)} 
\nonumber\\ \fl\qquad 
= {\hat{\mathcal S}}^{\dag} 
T_+
\exp\ensuremath{\big(
i\eta_+\hat A+i\zeta_+\hat J
 \big)} 
\exp\ensuremath{\bigg[
\frac{i}{\hbar }\ensuremath{\big(
\hat A\hat J+ A_{\mathrm{e}}\hat J+\hat A J_{\mathrm{e}}
 \big)}
\bigg]}, 
\label{eq:8VX} % \nonumber % \Z 
\\
\fl T_-
\exp\ensuremath{\big(
-i\eta_-{\hat{\mathcal A}}-i\zeta_-{\hat{\mathcal J}}
 \big)} 
\nonumber\\ \fl\qquad 
= 
\ensuremath{\bigg\{
T_-
\exp\ensuremath{\big(
-i\eta_-\hat A-i\zeta_-\hat J
 \big)} 
%\\ \times 
\exp\ensuremath{\bigg[
-\frac{i}{\hbar }\ensuremath{\big(
\hat A\hat J+ A_{\mathrm{e}}\hat J+\hat A J_{\mathrm{e}}
 \big)}
\bigg]}
\bigg\}} {\hat{\mathcal S}}, 
\label{eq:51ZF} % \nonumber % \Z 
\end{eqnarray}}%
%+++++++++++++++++++++++++++++++++++++++++++++
Unlike in equations (\ref{eq:93HW}), (\ref{eq:95HY}), in equations (\ref{eq:6XH})--(\ref{eq:51ZF}) the distinction between operators ${\hat{\mathcal S}}$ and ${\hat{\mathcal S}}^{\dag}$ and their representation as time-ordered operator functionals is explicitly maintained. When substituting equations (\ref{eq:8VX}), (\ref{eq:51ZF}) into (\ref{eq:2XC}), the ${\hat{\mathcal S}}^{\dag}$ and ${\hat{\mathcal S}}$ factors cancel each other, 
%=============================================
{\begin{eqnarray}\eqalign{ 
{\hat{\mathcal S}}{\hat{\mathcal S}}^{\dag} = \hat 1, 
}%
\label{eq:57ZN} % \nonumber % \Z 
\end{eqnarray}}%
%+++++++++++++++++++++++++++++++++++++++++++++
and we have, 
%=============================================
{\begin{eqnarray} 
\fl \Xi\ensuremath{\big(
{\eta}_+,{\eta}_-,{\zeta}_+,{\zeta}_-\big| J_{\mathrm{e}} ,A_{\mathrm{e}}
 \big)} 
\nonumber\\ \fl\qquad 
= \ensuremath{\bigg\langle
T_-
\exp\ensuremath{\big(
-i\eta_-\hat A-i\zeta_-\hat J
 \big)} 
\exp\ensuremath{\bigg[
-\frac{i}{\hbar }\ensuremath{\big(
\hat A\hat J+ A_{\mathrm{e}}\hat J+\hat A J_{\mathrm{e}}
 \big)}
\bigg]}
\nonumber\\ \fl\qquad \qquad\times 
T_+
\exp\ensuremath{\big(
i\eta_+\hat A+i\zeta_+\hat J
 \big)} 
\exp\ensuremath{\bigg[
\frac{i}{\hbar }\ensuremath{\big(
\hat A\hat J+ A_{\mathrm{e}}\hat J+\hat A J_{\mathrm{e}}
 \big)}
\bigg]}\bigg\rangle} . 
\label{eq:4XE} % \nonumber % \Z 
\end{eqnarray}}%
%+++++++++++++++++++++++++++++++++++++++++++++
The same in terms of $T_C$-ordering reads, 
%\begin{widetext} 
%=============================================
{\begin{eqnarray} 
\fl \Xi\ensuremath{\big(
{\eta}_+,{\eta}_-,{\zeta}_+,{\zeta}_-\big| J_{\mathrm{e}} ,A_{\mathrm{e}}
 \big)} %\nfp 
= \ensuremath{\bigg\langle
T_C
\exp\ensuremath{\big(
i\eta_+\hat A_{+}
-i\eta_-\hat A_{-}
+i\zeta_+\hat J_{+}
-i\zeta_-\hat J_{-}
 \big)} 
\nonumber\\ \fl\qquad\times 
\exp\ensuremath{\bigg[
\frac{i}{\hbar }\ensuremath{\big(
\hat A_+\hat J_+ + A_{\mathrm{e}}\hat J_+ +\hat A_+ J_{\mathrm{e}}
 \big)}
\bigg]}
%\\ \times 
\exp\ensuremath{\bigg[
-\frac{i}{\hbar }\ensuremath{\big(
\hat A_-\hat J_- + A_{\mathrm{e}}\hat J_- +\hat A_- J_{\mathrm{e}}
 \big)}
\bigg]}
\bigg\rangle} . 
\label{eq:5XF} % \nonumber % \Z 
\end{eqnarray}}%
%+++++++++++++++++++++++++++++++++++++++++++++
%\end{widetext}%
Grouping of arguments in this formula is an allusion to its ``ancestors:'' equations (\ref{eq:2XC}), (\ref{eq:6XH}) and (\ref{eq:56ZM}). 

%********************************************************
\subsection{Wick's theorem and elimination of field operators}%
\label{ch:WEF}
%********************************************************
In the standard techniques of quantum field theory \cite{Schweber,VasF}, the next step is applying Wick's theorem \cite{Wick} to equation (\ref{eq:5XF}). Wick's theorem allows one to rewrite time-ordered operator structures in a normally-ordered form (for definitions of operator orderings see section \ref{ch:ADO}). This comes especially handy if the initial state of the system is vacuum. 

The interaction-picture operators $\hat A(x,t)$ and $\hat J(x,t)$ are defined in two orthogonal subspaces and may be manipulated independently. Due to assumed factorisation of the density matrix, this equally applies to quantum averaging. We therefore apply Wick's theorem only to the field operators in (\ref{eq:5XF}), keeping the current operators ``as are.'' 

We employ the functional (Hori's \cite{Hori}) form of Wick's theorem (the Hori-Wick theorem, for short). It is discussed by Vasil'ev \cite{VasF} and in our paper \cite{WickCaus}. The directly applicable case is that of real field, see paper \cite{WickCaus}, section VIII. The Hori-Wick theorem for a real field reads, 
%=============================================
{\begin{eqnarray}\eqalign{ 
 T_C\mathcal{F}(\hat A_{+},\hat A_{-}) = {\mbox{\rm\boldmath$:$}}
\mathcal{F}_N(\hat A ,\hat A )
{\mbox{\rm\boldmath$:$}} , 
\\ 
 \mathcal{F}_N( A_+, A_-) 
= \exp \mathcal{Z}_C\ensuremath{\bigg(
\frac{\delta }{\delta A_+},
\frac{\delta }{\delta A_-}
 \bigg)}
\mathcal{F}( A_+, A_-) , 
}%
\label{eq:64ZV} % \nonumber % \Z 
\end{eqnarray}}%
%+++++++++++++++++++++++++++++++++++++++++++++
where $ A_{\pm}(x,t)$ are auxiliary c-number functional arguments, and $\mathcal{F}(\cdot,\cdot)$ is an arbitrary c-number functional. 
The {\em reordering form\/} $\mathcal{Z}_C(\cdot,\cdot)$ is a functional bilinear form where the kernels are contractions, (with $f_{\pm}(x,t)$ being auxiliary functional arguments)
%=============================================
{\begin{eqnarray}\eqalign{ 
\mathcal{Z}_C\ensuremath{\big(
 f_+ ,
 f_- 
 \big)} = -\frac{i\hbar}{2}f_+\Delta _{\mathrm{F}} f_+ +\frac{i\hbar}{2}f_-\Delta_{\mathrm{F}}^* f_- 
%\\ 
-i\hbar f_-\Delta^{(+)} f_+ . 
}%
\label{eq:42UP} % \nonumber % \Z 
\end{eqnarray}}%
%+++++++++++++++++++++++++++++++++++++++++++++
We use notation (\ref{eq:76NU}). Differentiations are carried out using the relations, 
%=============================================
{\begin{eqnarray}\eqalign{ 
\frac{\delta A_+(x,t)}{\delta A_+(x',t')} = \frac{\delta A_-(x,t)}{\delta A_-(x',t')} = \delta (x-x')\delta (t-t') , 
\\
\frac{\delta A_+(x,t)}{\delta A_-(x',t')} = \frac{\delta A_-(x,t)}{\delta A_+(x',t')} = 0. 
}%
\label{eq:69AA} % \nonumber % \Z 
\end{eqnarray}}%
%+++++++++++++++++++++++++++++++++++++++++++++
Note that under the normal ordering in equation (\ref{eq:64ZV}) the ${}_{\pm}$ indices of the operators are dropped as irrelevant. 

For any functional $\mathcal{F}_N(\cdot,\cdot)$, 
%=============================================
{\begin{eqnarray}\eqalign{ 
\ensuremath{\big\langle 0\big|
{\mbox{\rm\boldmath$:$}}
\mathcal{F}_N(\hat A ,\hat A )
{\mbox{\rm\boldmath$:$}} 
\big|0\big\rangle} 
= \mathcal{F}_N(0,0) , 
}%
\label{eq:9VY} % \nonumber % \Z 
\end{eqnarray}}%
%+++++++++++++++++++++++++++++++++++++++++++++
and we find, 
%\begin{widetext} 
%=============================================
{\begin{eqnarray} 
\fl\Xi\ensuremath{\big(
{\eta}_+,{\eta}_-,{\zeta}_+,{\zeta}_-\big| J_{\mathrm{e}} ,A_{\mathrm{e}}
 \big)} 
= \exp \mathcal{Z}_C\ensuremath{\bigg(
\frac{\delta }{\delta A_+},
\frac{\delta }{\delta A_-}
 \bigg)}
\nonumber\\ \fl\qquad \times 
\exp\ensuremath{\bigg[
i\ensuremath{\bigg(
\eta_+ + \frac{J_{\mathrm{e}}}{\hbar }
 \bigg)} A_+ -i\ensuremath{\bigg(
\eta_- + \frac{J_{\mathrm{e}}}{\hbar }
 \bigg)} A_-
\bigg]}
\nonumber\\ \fl\qquad \times 
\mathrm{Tr} \hat\rho_{\mathrm{dev}}T_C
\exp\ensuremath{\bigg[
i\hat J_{+}\ensuremath{\bigg(
\frac{ A_+ + A_{\mathrm{e}}}{\hbar }+\zeta_+
 \bigg)} 
%\\ 
-i\hat J_{-}\ensuremath{\bigg(
\frac{ A_- + A_{\mathrm{e}}}{\hbar }+\zeta_-
 \bigg)} 
\bigg]} 
\settoheight{\auxlv}{$\Big|$}%
\raisebox{-0.3\auxlv}{$\Big|_{ A_{\pm}=0}$}. 
\label{eq:84AS} % \nonumber % \Z 
\end{eqnarray}}%
%+++++++++++++++++++++++++++++++++++++++++++++
%\end{widetext}%
The necessary and sufficient information about the device is thus collected in the functional, 
%=============================================
{\begin{eqnarray}\eqalign{ 
\Xi_{\mathrm{dev}}^{\mathrm{I}}\ensuremath{\big(
\zeta_+,\zeta_-
 \big)} = \mathrm{Tr} \hat\rho_{\mathrm{dev}}T_C
\exp\ensuremath{\big(
i\zeta_+\hat J_{+}
-i\zeta_-\hat J_{-}
 \big)} 
.
}%
\label{eq:85AT} % \nonumber % \Z 
\end{eqnarray}}%
%+++++++++++++++++++++++++++++++++++++++++++++
With it we have, 
%=============================================
{\begin{eqnarray} 
\Xi\ensuremath{\big(
{\eta}_+,{\eta}_-,{\zeta}_+,{\zeta}_-\big| J_{\mathrm{e}} ,A_{\mathrm{e}}
 \big)} 
= \exp \mathcal{Z}_C\ensuremath{\bigg(
\frac{\delta }{\delta A_+},
\frac{\delta }{\delta A_-}
 \bigg)}
\nonumber\\ \qquad \times 
\exp\ensuremath{\bigg[
i\ensuremath{\bigg(
\eta_+ + \frac{J_{\mathrm{e}}}{\hbar }
 \bigg)} A_+ -i\ensuremath{\bigg(
\eta_- + \frac{J_{\mathrm{e}}}{\hbar }
 \bigg)} A_-
\bigg]}
\nonumber\\ \qquad \times 
\Xi_{\mathrm{dev}}^{\mathrm{I}}\ensuremath{\bigg(
\frac{ A_+ + A_{\mathrm{e}}}{\hbar }+\zeta_+,
\frac{ A_- + A_{\mathrm{e}}}{\hbar }+\zeta_-
 \bigg)}
\settoheight{\auxlv}{$\Big|$}%
\raisebox{-0.3\auxlv}{$\Big|_{ A_{\pm}=0}$}. 
\label{eq:86AU} % \nonumber % \Z 
\end{eqnarray}}%
%+++++++++++++++++++++++++++++++++++++++++++++

%********************************************************
\subsection{Response transformations revisited}%
\label{ch:IR}
%********************************************************
We are interested not in functional $\Xi $ as such but in its {\em response transformation\/} \cite{API,APII,APIII}, (cf.\ the closing remark in section \ref{ch:IC})
%=============================================
{\begin{eqnarray}\eqalign{ 
\Phi\ensuremath{\big(
\eta ,\zeta\big| j_{\mathrm{e}} , a_{\mathrm{e}}\big|
 J_{\mathrm{e}},A_{\mathrm{e}} \big)} %\\ 
= 
\Xi\ensuremath{\big(
{\eta}_+,{\eta}_-,{\zeta}_+,{\zeta}_-\big| J_{\mathrm{e}} ,A_{\mathrm{e}}
 \big)} \settoheight{\auxlv}{$|$}%
\raisebox{-0.3\auxlv}{$|_{\mathrm{c.v.}}$} 
, 
}%
\label{eq:13XQ} % \nonumber % \Z 
\end{eqnarray}}%
%+++++++++++++++++++++++++++++++++++++++++++++
where c.v.\ (short for {\em causal variables\/} \cite{WickCaus,BWO}) refers to the {\em response substitutions\/} \cite{API,APII,APIII,WickCaus}, %=============================================
{\begin{eqnarray} 
\eta_{\pm}(x,t) 
 = \frac{ j_{\mathrm{e}}(x,t)}{\hbar }\pm \eta ^{(\mp)}(x,t), 
\label{eq:5VU} % \nonumber % \Z 
\\ 
\zeta_{\pm}(x,t) 
= \frac{ a_{\mathrm{e}}(x,t)}{\hbar }\pm \zeta ^{(\mp)}(x,t). 
\label{eq:50JV} % \nonumber % \Z 
\end{eqnarray}}%
%+++++++++++++++++++++++++++++++++++++++++++++
The frequency-positive and negative\ parts are defined by equation (\ref{eq:4JH}). Formulae for the causal variables are, 
%=============================================
{\begin{eqnarray} 
\fl{\eqalign{
\eta (x,t) = \eta_{+}(x,t)-\eta_{-}(x,t), \qquad 
 j_{\mathrm{e}}(x,t) = \hbar \ensuremath{\big[
\eta_+ ^{(+)}(x,t)+\eta_- ^{(-)}(x,t)
\big]} ,
}} 
\label{eq:7VW} % \nonumber % \Z 
\\ 
\fl{\eqalign{
\zeta (x,t) = \zeta_{+}(x,t)-\zeta_{-}(x,t), \qquad 
 a_{\mathrm{e}}(x,t) = \hbar \ensuremath{\big[
\zeta_+ ^{(+)}(x,t)+\zeta_- ^{(-)}(x,t)
\big]} . 
}} 
\label{eq:6VV} % \nonumber % \Z 
\end{eqnarray}}%
%+++++++++++++++++++++++++++++++++++++++++++++
Equations (\ref{eq:5VU}), (\ref{eq:50JV}) and (\ref{eq:7VW}), (\ref{eq:6VV}) are mutually inverse, showing that these formulae constitute a genuine change of functional variables. Response substitutions are inherently related to equations (\ref{eq:11JQ}). For details see our paper \cite{API}. 
%********************************************************
\subsection{The causal Wick theorem revisited}%
\label{ch:ICW}
%********************************************************
For the two exponents in (\ref{eq:86AU}), the laws of transformation were worked out in paper \cite{WickCaus}, 
section VIIIB. 
With \mbox{$
\Xi_{\mathrm{dev}}^{\mathrm{I}}=1
$} and \mbox{$J_{\mathrm{e}}(x,t)=0$}, equation (\ref{eq:86AU}) reduces to the ``test-case formula'' 
(104) in paper \cite{WickCaus}. To rewrite that relation in causal variables we applied additional response substitution, 
%=============================================
{\begin{eqnarray}\eqalign{ 
 A_{\pm}(x,t) = a_{\mathrm{e}}'(x,t)\pm\hbar \zeta^{\prime(\mp)}(x,t). 
}%
\label{eq:38JH} % \nonumber % \Z 
\end{eqnarray}}%
%+++++++++++++++++++++++++++++++++++++++++++++
Variables $ a_{\mathrm{e}}'(x,t)$, 
$\zeta'(x,t)$ are given by the formulae, 
%=============================================
{\begin{eqnarray}\eqalign{ 
\fl\zeta '(x,t) = \frac{ A_{+}(x,t)- A_{-}(x,t)}{\hbar }, \qquad 
 a_{\mathrm{e}}'(x,t) = A_{+}^{(+)}(x,t)+ A_{-}^{(-)}(x,t) . 
}%
\label{eq:10VZ} % \nonumber % \Z 
\end{eqnarray}}%
%+++++++++++++++++++++++++++++++++++++++++++++
Equations (\ref{eq:38JH}), (\ref{eq:10VZ}) are a rescaled version of (\ref{eq:50JV}), (\ref{eq:6VV}): they turn into the latter if we define, 
%=============================================
{\begin{eqnarray}\eqalign{ 
\zeta_{\pm}'(x,t) = \frac{ A_{\pm}(x,t)}{\hbar } . 
}%
\label{eq:11WA} % \nonumber % \Z 
\end{eqnarray}}%
%+++++++++++++++++++++++++++++++++++++++++++++
Primes keep variables (\ref{eq:10VZ}) distinct from (\ref{eq:6VV}). 

In causal variables, 
%=============================================
{\begin{eqnarray}\eqalign{ 
\ensuremath{\bigg[
i\ensuremath{\bigg(
\eta_+ + \frac{J_{\mathrm{e}}}{\hbar }
 \bigg)} A_+ -i\ensuremath{\bigg(
\eta_- + \frac{J_{\mathrm{e}}}{\hbar }
 \bigg)} A_-
\bigg]}\settoheight{\auxlv}{$\Big|$}%
\raisebox{-0.3\auxlv}{$\Big|_{{\mathrm{c.v.}}}$} %\\ 
= \eta a_{\mathrm{e}}'+\zeta '\ensuremath{\big(
j_{\mathrm{e}}+J_{\mathrm{e}}
 \big)} , 
}%
\label{eq:41JL} % \nonumber % \Z 
\end{eqnarray}}%
%+++++++++++++++++++++++++++++++++++++++++++++
and 
%=============================================
{\begin{eqnarray}\eqalign{ 
\mathcal{Z}_C\ensuremath{\bigg(
\frac{\delta }{\delta A_+},
\frac{\delta }{\delta A_-}
 \bigg)}\settoheight{\auxlv}{$\Big|$}%
\raisebox{-0.3\auxlv}{$\Big|_{{\mathrm{c.v.}}}$} = - i 
\frac{\delta }{\delta a_{\mathrm{e}}'} \Delta_{\mathrm{R}} \frac{\delta }{\delta \zeta'}, 
}%
\label{eq:36KS} % \nonumber % \Z 
\end{eqnarray}}%
%+++++++++++++++++++++++++++++++++++++++++++++
where we use condensed notation (\ref{eq:3VS}), (\ref{eq:76NU}). 

%********************************************************
\subsection{Closed perturbative formula in causal variables}%
\label{ch:DPC}
%********************************************************
As to the functional $\Xi_{\mathrm{dev}}^{\mathrm{I}} $, we define, 
%=============================================
{\begin{eqnarray}\eqalign{ 
\Phi_{\mathrm{dev}}^{\mathrm{I}}\ensuremath{\big(
\zeta\big|a_{\mathrm{e}} 
 \big)} = 
\Xi_{\mathrm{dev}}^{\mathrm{I}} ({\zeta}_+,{\zeta}_-)\settoheight{\auxlv}{$|$}%
\raisebox{-0.3\auxlv}{$|_{\mathrm{c.v.}}$} , 
}%
\label{eq:17XU} % \nonumber % \Z 
\end{eqnarray}}%
%+++++++++++++++++++++++++++++++++++++++++++++
where c.v.\ refers to substitution (\ref{eq:6VV}). Combining (\ref{eq:6VV}) and (\ref{eq:38JH}) it is easy to show that, 
%=============================================
{\begin{eqnarray} 
\frac{ A_{\mathrm{e}}(x,t) + A_{\pm}(x,t)}{\hbar }+\zeta_{\pm}(x,t) 
\nonumber\\ \qquad = 
\frac{ A_{\mathrm{e}}(x,t) + a_{\mathrm{e}}(x,t) + a_{\mathrm{e}}'(x,t)}{\hbar } 
%\\ 
\pm \zeta ^{(\mp)}(x,t)\pm\zeta^{'(\mp)}(x,t),
\label{eq:88AW} % \nonumber % \Z 
\end{eqnarray}}%
%+++++++++++++++++++++++++++++++++++++++++++++
hence 
%=============================================
{\begin{eqnarray}\eqalign{ \fl
\Xi_{\mathrm{dev}}^{\mathrm{I}}\ensuremath{\bigg(
\frac{ A_{\mathrm{e}} + A_+}{\hbar }+\zeta_+,
\frac{ A_{\mathrm{e}} + A_-}{\hbar }+\zeta_-
 \bigg)}\settoheight{\auxlv}{$\Big|$}%
\raisebox{-0.3\auxlv}{$\Big|_{\mathrm{c.v.}}$} %\\ 
= \Phi_{\mathrm{dev}}^{\mathrm{I}}
\ensuremath{\big(
\zeta +\zeta '\big|
 A_{\mathrm{e}}+ a_{\mathrm{e}} + a_{\mathrm{e}}'
 \big)} . 
}%
\label{eq:89AX} % \nonumber % \Z 
\end{eqnarray}}%
%+++++++++++++++++++++++++++++++++++++++++++++
Putting equations (\ref{eq:41JL}), (\ref{eq:36KS}) and (\ref{eq:89AX}) together we obtain the relation sought, 
%=============================================
{\begin{eqnarray} 
\fl \Phi\ensuremath{\big(
\eta ,\zeta\big| j_{\mathrm{e}} , a_{\mathrm{e}}\big|
 J_{\mathrm{e}},A_{\mathrm{e}} \big)} 
= 
\exp\ensuremath{\bigg(
- i 
\frac{\delta }{\delta a_{\mathrm{e}}'} \Delta_{\mathrm{R}} 
\frac{\delta }{\delta \zeta '}
 \bigg)} 
\nonumber\\ \fl\qquad \times 
\exp\ensuremath{\big[
i\eta a_{\mathrm{e}}' 
+i\zeta'\ensuremath{\big(
j_{\mathrm{e}}+J_{\mathrm{e}}
 \big)} 
\big]} 
\Phi_{\mathrm{dev}}^{\mathrm{I}}\ensuremath{\big(
\zeta+\zeta ' \big| A_{\mathrm{e}}+ a_{\mathrm{e}}+ a_{\mathrm{e}}'
 \big)} \settoheight{\auxlv}{$|$}%
\raisebox{-0.3\auxlv}{$|_{\zeta '= a_{\mathrm{e}}'=0}$} . 
\label{eq:19XW} % \nonumber % \Z 
\end{eqnarray}}%
%+++++++++++++++++++++++++++++++++++++++++++++
The generalised consistency condition \cite{APII,APIII}, 
%=============================================
{\begin{eqnarray} \fl
\Phi\ensuremath{\big(
\eta ,\zeta\big| j_{\mathrm{e}} , a_{\mathrm{e}}\big|
 J_{\mathrm{e}},A_{\mathrm{e}} \big)} %\\ 
& = \Phi\ensuremath{\big(
\eta ,\zeta\big| j_{\mathrm{e}}+J_{\mathrm{e}} , 
a_{\mathrm{e}}+A_{\mathrm{e}} \big|
0,0 \big)} 
%\nonumber\\ & 
\equiv \Phi\ensuremath{\big(
\eta ,\zeta\big| j_{\mathrm{e}}+J_{\mathrm{e}} , 
a_{\mathrm{e}}+A_{\mathrm{e}} \big)}, 
\label{eq:52AU} % \nonumber % \Z 
\end{eqnarray}}%
%+++++++++++++++++++++++++++++++++++++++++++++
is evident in (\ref{eq:19XW}). 

In what follows, we set $A_{\mathrm{e}}(x,t)$ and $J_{\mathrm{e}}(x,t)$ to zero. Relations with sources, which are important for physics, may always be restored replacing $a_{\mathrm{e}}(x,t)\to a_{\mathrm{e}}(x,t)+A_{\mathrm{e}}(x,t)$ and $j_{\mathrm{e}}(x,t)\to j_{\mathrm{e}}(x,t)+J_{\mathrm{e}}(x,t)$. 
%********************************************************
\subsection{Dressing the current}%
\label{ch:DC}
%********************************************************
Unlike in the linear case \cite{WickCaus}, finding a closed solution to (\ref{eq:19XW}) is certainly impossible. One can, however, reduce (\ref{eq:19XW}) to the functional, 
%=============================================
{\begin{eqnarray}\eqalign{ 
\Phi_{\mathrm{dev}}\ensuremath{\big(
\zeta\big|a_{\mathrm{e}} \big)} 
= \Phi\ensuremath{\big(
0 ,\zeta\big| 0 , a_{\mathrm{e}} \big)} = \Phi\ensuremath{\big(
0 ,\zeta\big| 0 , a_{\mathrm{e}}\big| 0 ,0 \big)} , 
}%
\label{eq:20XX} % \nonumber % \Z 
\end{eqnarray}}%
%+++++++++++++++++++++++++++++++++++++++++++++
cf.\ equation (\ref{eq:52AU}). This functional expresses the properties of the {Heisenberg}\ current operator (physical, or ``dressed,'' current). A perturbative formula for $\Phi_{\mathrm{dev}}$ is given by equation (\ref{eq:19XW}) with $\eta (x,t) = j_{\mathrm{e}}(x,t) = A_{\mathrm{e}}(x,t) = J_{\mathrm{e}}(x,t) =0$, 
%=============================================
{\begin{eqnarray}\eqalign{ 
\Phi_{\mathrm{dev}}\ensuremath{\big(
\zeta\big|a_{\mathrm{e}} \big)} 
= 
\exp\ensuremath{\bigg(
- i 
\frac{\delta }{\delta a_{\mathrm{e}}} \Delta_{\mathrm{R}} 
\frac{\delta }{\delta \zeta }
 \bigg)} 
\Phi_{\mathrm{dev}}^{\mathrm{I}}\ensuremath{\big(
\zeta\big|a_{\mathrm{e}} \big)} . 
}%
\label{eq:23YA} % \nonumber % \Z 
\end{eqnarray}}%
%+++++++++++++++++++++++++++++++++++++++++++++
Functional $\Phi_{\mathrm{dev}}^{\mathrm{I}}$ expresses the properties of the interaction-pictire current operator (``bare'' current). Equation (\ref{eq:23YA}) thus expresses ``dressing'' of the current operator by the electromagnetic interaction. 

Finding the dressed current is the hard part of the problem. Vast majority of technical developments in quantum field theory, condensed matter physics and quantum optics are in essense attempts to approximate equation (\ref{eq:23YA}) or its counterparts. Connections between equation (\ref{eq:23YA}) and diagram techniques are elucidated by Vasil'ev \cite{VasF}. In \ref{ch:Diag} we construct a formal solution to (\ref{eq:23YA}) in terms of a diagram expansion. That (\ref{eq:23YA}) may be used as an alternative entry point to phase-space techniques was shown in paper \cite{BWO}. 
%********************************************************
\subsection{Solution in terms of dressed current}%
\label{ch:SDC}
%********************************************************
Assuming that equation (\ref{eq:23YA}) is solved, full electromagnetic properties of the device follow with ease. We employ the formula, 
%=============================================
{\begin{eqnarray}\eqalign{ 
\exp\ensuremath{\bigg(
f\frac{\delta }{\delta g}
 \bigg)} 
\exp\ensuremath{\big(
igh
 \big)} %\\ 
= \exp\ensuremath{\big(
igh
 \big)}\exp\ensuremath{\bigg[
f\ensuremath{\bigg(\frac{\delta }{\delta g}+ih
 \bigg)}\bigg]} , 
}%
\label{eq:25YC} % \nonumber % \Z 
\end{eqnarray}}%
%+++++++++++++++++++++++++++++++++++++++++++++
where $f(x,t)$, $g(x,t)$ and $h(x,t)$ are c-number functions. 
It can be verified, e.g., expanding all exponents in Taylor series. With it we can pull $\exp\ensuremath{\big(
i\eta a_{\mathrm{e}}' 
+i\zeta' j_{\mathrm{e}}
 \big)}$ from under the differentiation resulting in, 
%=============================================
{\begin{eqnarray} \fl
\Phi\ensuremath{\big(
\eta ,\zeta \big| j_{\mathrm{e}}, a_{\mathrm{e}} 
 \big)} 
%\\ 
= 
\exp\ensuremath{\big(
i\eta a_{\mathrm{e}}' 
+i\zeta' j_{\mathrm{e}}
 \big)} 
\nonumber\\ \fl\qquad \times 
\exp\ensuremath{\bigg[
- i 
\ensuremath{\bigg(
\frac{\delta }{\delta a_{\mathrm{e}}'}+i\eta
 \bigg)} 
 \Delta_{\mathrm{R}} 
\ensuremath{\bigg(
\frac{\delta }{\delta \zeta '}+i j_{\mathrm{e}} 
 \bigg)}
\bigg]} 
%\\ \times 
\Phi_{\mathrm{dev}}^{\mathrm{I}}\ensuremath{\big(
\zeta+\zeta ' \big| a_{\mathrm{e}}+ a_{\mathrm{e}}'
 \big)} \settoheight{\auxlv}{$|$}%
\raisebox{-0.3\auxlv}{$|_{\zeta '= a_{\mathrm{e}}'=0}$} . 
\label{eq:20WL} % \nonumber % \Z 
\end{eqnarray}}%
%+++++++++++++++++++++++++++++++++++++++++++++
Setting $\zeta '(x,t)$ and $ a_{\mathrm{e}}'(x,t)$ to zero we find, 
%=============================================
{\begin{eqnarray}\eqalign{ \fl
\Phi\ensuremath{\big(
\eta ,\zeta \big| j_{\mathrm{e}}, a_{\mathrm{e}} 
 \big)} 
%\\ 
= \exp\ensuremath{\bigg[
- i 
\ensuremath{\bigg(
\frac{\delta }{\delta a_{\mathrm{e}}}+i\eta
 \bigg)} 
 \Delta_{\mathrm{R}} 
\ensuremath{\bigg(
\frac{\delta }{\delta \zeta }+i j_{\mathrm{e}} 
 \bigg)}
\bigg]} 
%\\ \times 
\Phi_{\mathrm{dev}}^{\mathrm{I}}\ensuremath{\big(
\zeta\big| a_{\mathrm{e}}
 \big)} . 
}%
\label{eq:26YD} % \nonumber % \Z 
\end{eqnarray}}%
%+++++++++++++++++++++++++++++++++++++++++++++
Bilinearity of the form in the exponent allows us to write, 
%=============================================
{\begin{eqnarray} \fl
\exp\ensuremath{\bigg[
- i 
\ensuremath{\bigg(
\frac{\delta }{\delta a_{\mathrm{e}}}+i\eta
 \bigg)} 
 \Delta_{\mathrm{R}} 
\ensuremath{\bigg(
\frac{\delta }{\delta \zeta }+i j_{\mathrm{e}} 
 \bigg)} \bigg]} \nonumber\\ \fl\qquad 
= \exp\ensuremath{\big(
i\eta\Delta_{\mathrm{R}} j_{\mathrm{e}}
 \big)} 
%\\ \times 
\exp\ensuremath{\bigg(
\frac{\delta }{\delta a_{\mathrm{e}}}\Delta_{\mathrm{R}} j_{\mathrm{e}}
 \bigg)} 
\exp\ensuremath{\bigg(
\eta\Delta_{\mathrm{R}} \frac{\delta }{\delta \zeta }
 \bigg)} 
%\\ \times 
\exp\ensuremath{\bigg(
-i 
\frac{\delta }{\delta a_{\mathrm{e}}} 
\Delta _{\mathrm{R}} 
\frac{\delta }{\delta \zeta }
 \bigg)} . 
\label{eq:27YE} % \nonumber % \Z 
\end{eqnarray}}%
%+++++++++++++++++++++++++++++++++++++++++++++
The last factor here is the dressing operator, cf.\ equation (\ref{eq:23YA}). The second and the third ones are functional shift operators. Applying the dressing and the two shifts to $\Phi_{\mathrm{dev}}^{\mathrm{I}}$ we obtain, 
%=============================================
{\begin{eqnarray} \fl 
\Phi\ensuremath{\big(
\eta , \zeta \big|j_{\mathrm{e}}, a_{\mathrm{e}}\big| J_{\mathrm{e}}, A_{\mathrm{e}} 
 \big)} \nonumber\\ \fl\qquad 
= \exp\ensuremath{\big[
i\eta\Delta_{\mathrm{R}} \ensuremath{\big(
j_{\mathrm{e}} + J_{\mathrm{e}}
 \big)} 
\big]} 
%\\ \times 
\Phi_{\mathrm{dev}}\ensuremath{\big(
\zeta+\eta\Delta_{\mathrm{R}}\big| a_{\mathrm{e}}+A_{\mathrm{e}}+ \Delta_{\mathrm{R}}\ensuremath{\big(
j_{\mathrm{e}} + J_{\mathrm{e}}
 \big)}
 \big)} , 
\label{eq:28YF} % \nonumber % \Z 
\end{eqnarray}}%
%+++++++++++++++++++++++++++++++++++++++++++++
where we use abbreviated notation of equations (\ref{eq:77NV}) and (\ref{eq:78NW}). 
In equation (\ref{eq:28YF}) the external sources are restored. 
Without the device ($\Phi_{\mathrm{dev}}=1$) and with $J_{\mathrm{e}}(x,t)=0$ we recover the test-case 
formula of paper \cite{WickCaus}, 
%=============================================
{\begin{eqnarray}\eqalign{ 
\Phi_{\mathrm{vac}}\ensuremath{\big(
\eta \big| j_{\mathrm{e}}
 \big)} = \exp\ensuremath{\big(
i\eta\Delta_{\mathrm{R}} j_{\mathrm{e}}
 \big)} . 
}%
\label{eq:15WE} % \nonumber % \Z 
\end{eqnarray}}%
%+++++++++++++++++++++++++++++++++++++++++++++
The physical content of equations (\ref{eq:23YA}) and (\ref{eq:28YF}) is a subject of forthcoming papers. %********************************************************
%********************************************************
\section{Oscillators interacting with quantised dipole (the narrow-band\ case)}%
\label{ch:EII}
%********************************************************
\subsection{Closed-time-loop formalism and perturbation theory}%
\label{ch:CTP}
%********************************************************
We now consider the second generic case: a quantised dipole interacting with a set of oscillators under the RWA. Formally, one sets $M=N$ in equations (\ref{eq:4SX}), (\ref{eq:45LB}). To a large extent, amendments to the broad-band\ case\ reduce to considering 
two fields, ${\hat{\mathcal A}}(x,t)\to{\hat{\mathcal E}}(x,t),{\hat{\mathcal E}}^{\dag}(x,t)$, 
and two ``currents,'' ${\hat{\mathcal J}}(x,t)\to{\hat{\mathcal D}}(x,t),{\hat{\mathcal D}}^{\dag}(x,t)$, 
with ${\hat{\mathcal E}}(x,t)$ coupled to ${\hat{\mathcal D}}^{\dag}(x,t)$ and ${\hat{\mathcal E}}^{\dag}(x,t)$ 
to ${\hat{\mathcal D}}(x,t)$. We remind that ${\hat{\mathcal E}}(x,t)$, ${\hat{\mathcal D}}(x,t)$ are Heisenberg counterparts of the interaction-picture operators $\hat E (x,t)$, $\hat D (x,t)$; $\hat E (x,t)$ is given by equation (\ref{eq:4SX}) and $\hat D (x,t)$ is ``just assumed known.'' 
Consequently all arguments in characteristic 
functionals also double, 
$\eta_{\pm}(x,t)\to\mu_{\pm}(x,t),\bar\mu_{\pm}(x,t)$, 
$\zeta_{\pm}(x,t)\to\nu_{\pm}(x,t),\bar\nu_{\pm}(x,t)$, etc. 
The basic object of the theory, 
the characteristic functional of the $T_C$-ordered products 
of the Heisenberg field and dipole operators, reads, (cf.\ the closing remark in section \ref{ch:IC})
%\pctxt
%=============================================
{\begin{eqnarray} 
\Xi\ensuremath{\big(
\bar\mu_+ ,\mu_+, 
\bar\mu_- ,\mu_-, 
\bar\nu_+ ,\nu_+, 
\bar\nu_- ,\nu_-
\big|D_{\mathrm{e}},D_{\mathrm{e}}^*,E_{\mathrm{e}},E_{\mathrm{e}}^* 
 \big)} 
\nonumber\\ \qquad 
= \ensuremath{\Big\langle T_C 
\exp\ensuremath{\big(
i\bar\mu_+{\hat{\mathcal E}}_+ +i\mu_+{\hat{\mathcal E}}_+^{\dag}
-i\bar\mu_-{\hat{\mathcal E}}_- -i\mu_-{\hat{\mathcal E}}_-^{\dag} 
\nonumber\\ \qquad \qquad
+i\bar\nu_+{\hat{\mathcal D}}_++i\nu_+{\hat{\mathcal D}}_+^{\dag}
-i\bar\nu_-{\hat{\mathcal D}}_- -i\nu_-{\hat{\mathcal D}}_-^{\dag}
 \big)} 
\Big\rangle} . 
\label{eq:33YM} % \nonumber % \Z 
\end{eqnarray}}%
%+++++++++++++++++++++++++++++++++++++++++++++
We again employ condensed notation (\ref{eq:3VS}). The S-matrix and its adjoint now are, 
%=============================================
{\begin{eqnarray} \fl
{\hat{\mathcal S}} = T_+\exp\ensuremath{\bigg\{
\frac{i}{\hbar }\ensuremath{\big[
\hat E \hat D^{\dag} +E_{\mathrm{e}}\hat D ^{\dag}+\hat E\hat D_{\mathrm{e}}^* 
%\\ 
+\hat E^{\dag} \hat D+E_{\mathrm{e}}^*\hat D+\hat E^{\dag}\hat D_{\mathrm{e}} 
\big]}
\bigg\}} , 
\label{eq:30YJ} % \nonumber % \Z 
\\ \fl 
{\hat{\mathcal S}}^{\dag} = T_-\exp\ensuremath{\bigg\{
-\frac{i}{\hbar }\ensuremath{\big[
\hat E \hat D^{\dag} +E_{\mathrm{e}}\hat D ^{\dag}+\hat E\hat D_{\mathrm{e}}^* 
%\\ 
+\hat E^{\dag} \hat D+E_{\mathrm{e}}^*\hat D+\hat E^{\dag}\hat D_{\mathrm{e}} 
\big]}
\bigg\}} , 
\label{eq:31YK} % \nonumber % \Z 
\end{eqnarray}}%
%+++++++++++++++++++++++++++++++++++++++++++++
By the same means as equation (\ref{eq:5XF}) was obtained we find a perturbative formula, 
%\begin{widetext} 
%=============================================
{\begin{eqnarray} \fl 
\Xi\ensuremath{\big(
\bar\mu_+ ,\mu_+, 
\bar\mu_- ,\mu_-, 
\bar\nu_+ ,\nu_+, 
\bar\nu_- ,\nu_-
\big|D_{\mathrm{e}},D_{\mathrm{e}}^*,E_{\mathrm{e}},E_{\mathrm{e}}^* 
 \big)} 
\nonumber\\ \fl\qquad 
= \ensuremath{\bigg\langle 
T_C 
\exp\ensuremath{\big(
i\bar\mu_+\hat E_{+}+i\mu_+\hat E_{+}^{\dag}
-i\bar\mu_-\hat E_{-} -i\mu_-\hat E_{-}^{\dag}
\nonumber\\ \fl\qquad\qquad\qquad\qquad\qquad 
+ i\bar\nu_+\hat D_{+}+i\nu_+\hat D_{+}^{\dag}
-i\bar\nu_-\hat D_{-} -i\nu_-\hat D_{-}^{\dag}
 \big)} 
\nonumber\\ \fl\qquad\quad\times 
\exp\ensuremath{\bigg\{
\frac{i}{\hbar }\ensuremath{\big[
\hat E_{+} \hat D^{\dag}_{+} +E_{\mathrm{e}}\hat D_{+} ^{\dag}+\hat E_{+}\hat D_{\mathrm{e}}^* 
+\hat E_{+}^{\dag} \hat D_{+}+E_{\mathrm{e}}^*\hat D_{+}+\hat E_{+}^{\dag}\hat D_{\mathrm{e}} 
\big]}
\bigg\}}
\nonumber\\ \fl\qquad\quad\times 
\exp\ensuremath{\bigg\{
-\frac{i}{\hbar }\ensuremath{\big[
\hat E_{-} \hat D^{\dag}_{-} +E_{\mathrm{e}}\hat D_{-} ^{\dag}+\hat E_{-}\hat D_{\mathrm{e}}^* 
+\hat E_{-}^{\dag} \hat D_{-}+E_{\mathrm{e}}^*\hat D_{-}+\hat E_{-}^{\dag}\hat D_{\mathrm{e}} 
\big]}
\bigg\}}
\bigg\rangle} . 
\label{eq:34YN} % \nonumber % \Z 
\end{eqnarray}}%
%+++++++++++++++++++++++++++++++++++++++++++++
The applicable version of the Hori-Wick theorem is that for the complex field in the nonrelativistic case 
(cf.\ paper \cite{WickCaus}, 
eq.\ (23)): 
%=============================================
{\begin{eqnarray}\eqalign{ \fl
 T_C\mathcal{F}(\hat E_{+},\hat E_{+}^{\dag},\hat E_{-},\hat E_{-}^{\dag}) 
= {\mbox{\rm\boldmath$:$}}\mathcal{F}_N(\hat E_{},\hat E_{}^{\dag}, 
\hat E_{},\hat E_{}^{\dag})
{\mbox{\rm\boldmath$:$}} , 
\\ \fl 
\mathcal{F}_N( E_+,\bar E_+ ,E_-, \bar E_-) 
%\\ 
= \exp Z_C\ensuremath{\bigg(
\frac{\delta }{\delta E_+},
\frac{\delta }{\delta \bar E_+},
\frac{\delta }{\delta E_-},
\frac{\delta }{\delta \bar E_-}
 \bigg)}
%\\ \times 
\mathcal{F}( E_+, \bar E_+ ,E_-, \bar E_-), 
}%
\label{eq:38UK} % \nonumber % \Z 
\end{eqnarray}}%
%+++++++++++++++++++++++++++++++++++++++++++++
where $E_{\pm}(x,t),\bar E_{\pm}(x,t)$ are four auxiliary functional arguments. The reordering form $Z_C$ reads, (in notation (\ref{eq:76NU}))
%=============================================
{\begin{eqnarray}\eqalign{ 
Z_C\ensuremath{\big(
\bar f_+,
f_-,
\bar f_-,
f_-
 \big)} %\\ 
= -i\hbar \bar f_+G_{\mathrm{F}} f_+
+i\hbar \bar f_- G_{\mathrm{F}}^* f_-
-i\hbar \bar f_- G^{(+)} f_+
. 
}%
\label{eq:40UM} % \nonumber % \Z 
\end{eqnarray}}%
%+++++++++++++++++++++++++++++++++++++++++++++
$f_{\pm}(x,t),\bar f_{\pm}(x,t)$ are auxiliary functional arguments. Using that 
%=============================================
{\begin{eqnarray}\eqalign{ 
\ensuremath{\big\langle 0\big|
{\mbox{\rm\boldmath$:$}}\mathcal{F}_N(\hat E_{},\hat E_{}^{\dag}, 
\hat E_{},\hat E_{}^{\dag})
{\mbox{\rm\boldmath$:$}}
\big|0\big\rangle} = \mathcal{F}_N(0,0,0,0), 
}%
\label{eq:16WF} % \nonumber % \Z 
\end{eqnarray}}%
%+++++++++++++++++++++++++++++++++++++++++++++
we find the narrow-band\ counterpart of equation (\ref{eq:86AU}), 
%=============================================
{\begin{eqnarray} 
\fl\Xi\ensuremath{\big(
\bar\mu_+ ,\mu_+, 
\bar\mu_- ,\mu_-, 
\bar\nu_+ ,\nu_+, 
\bar\nu_- ,\nu_-
\big|D_{\mathrm{e}},D_{\mathrm{e}}^*,E_{\mathrm{e}},E_{\mathrm{e}}^* 
 \big)} 
= \exp Z_C\ensuremath{\bigg(
\frac{\delta }{\delta E_+},
\frac{\delta }{\delta \bar E_+},
\frac{\delta }{\delta E_-},
\frac{\delta }{\delta \bar E_-}
 \bigg)}
\nonumber\\ \fl\ \ \times 
\exp\ensuremath{\bigg[ 
i\ensuremath{\bigg(
\bar\mu_++\frac{D_{\mathrm{e}}^*}{\hbar }
 \bigg)}E_++i\ensuremath{\bigg(
\mu_++\frac{D_{\mathrm{e}}}{\hbar }
 \bigg)}\bar E_+ 
-i\ensuremath{\bigg(
\bar\mu_-+\frac{D_{\mathrm{e}}^*}{\hbar }
 \bigg)}E_--i\ensuremath{\bigg(
\mu_- +\frac{D_{\mathrm{e}}}{\hbar }
 \bigg)}\bar E_-
\bigg]}
\nonumber\\ \fl\ \ \times 
\Xi ^{\mathrm{dev}} 
\ensuremath{\bigg(
\bar\nu_+ + \frac{E_{\mathrm{e}}^*+\bar E_+}{\hbar}, 
\nu_+ + \frac{E_{\mathrm{e}}+E_+}{\hbar}, 
\bar\nu_- + \frac{E_{\mathrm{e}}^*+\bar E_-}{\hbar}, 
\nu_- + \frac{E_{\mathrm{e}}+E_-}{\hbar}
 \bigg)} 
\settoheight{\auxlv}{$\Big|$}%
\raisebox{-0.3\auxlv}{$\Big|_{E_{\pm}=\bar E_{\pm}=0}$}. \nonumber\\ \fl\qquad
\label{eq:37YR} % \nonumber % \Z 
\end{eqnarray}}%
%+++++++++++++++++++++++++++++++++++++++++++++
%\end{widetext}%
where 
%=============================================
{\begin{eqnarray}\eqalign{ \fl 
\Xi ^{\mathrm{dev}}\ensuremath{\big(
\bar\nu_+,\nu_+,\bar\nu_-,\nu_- 
 \big)} 
%\\ 
= \mathrm{Tr} \hat \rho_{\mathrm{dev}} T_C 
\exp\ensuremath{\big(
i\bar\nu_+ \hat D_{+}
+i\nu_+ \hat D_{+}^{\dag}
-i\bar\nu_- \hat D_{-}
-i\nu_- \hat D_{-}^{\dag}
 \big)} . 
}%
\label{eq:90AY} % \nonumber % \Z 
\end{eqnarray}}%
%+++++++++++++++++++++++++++++++++++++++++++++
%********************************************************
\subsection{Plain {\em versus\/} duplicate phase space}%
\label{ch:RC}
%********************************************************
In paper \cite{BWO}, techniques similar to those developed here were used to derive various generalisations of the positive-P representation of quantum optics (see \cite{BWO} and references therein). The positive-P as well as its generalisations assume doubling of the classical phase space. This doubling is also present in response transformations for the nonresonant\ case, derived in paper \cite{WickCaus}, 
section IV. 
They were deduced for the ``test-case formula'' which is nothing but equation (\ref{eq:37YR}) with $\Xi ^{\mathrm{dev}}=1$ and $D_{\mathrm{e}}(x,t)=0$ (i.e., for a free field in a vacuum state). Direct calculation (see paper \cite{WickCaus}, 
section VC) 
gives, 
%=============================================
{\begin{eqnarray}\eqalign{ 
\Xi_{\mathrm{vac}}\ensuremath{\big(
\bar\mu_+ ,\mu_+, 
\bar\mu_- ,\mu_- 
 \big)}
= \exp Z_C (i\bar\mu_+,i\mu_+,-i\bar\mu_-,-i\mu_-) . 
}%
\label{eq:44JP} % \nonumber % \Z 
\end{eqnarray}}%
%+++++++++++++++++++++++++++++++++++++++++++++
Response substitutions in the ``test-case formula'' read, 
%=============================================
{\begin{eqnarray}\eqalign{ 
\mu_+
%(x, \quad t) 
 = \frac{d_{\mathrm{e}}
%(x, \quad t) 
}{\hbar }, \quad 
\bar\mu_+
%(x, \quad t) 
 = \bar\mu
%(x, \quad t) 
+ \frac{\bar d_{\mathrm{e}}
%(x, \quad t) 
}{\hbar }, \quad 
\bar\mu_-
%(x, \quad t) 
 = \frac{\bar d_{\mathrm{e}}
%(x, \quad t) 
}{\hbar }, \quad 
\mu_-
%(x, \quad t) 
 = \mu
%(x, \quad t) 
 + \frac{d_{\mathrm{e}}
%(x, \quad t) 
}{\hbar }, 
\\ 
E_+
%(x, \quad t) 
 = e_{\mathrm{e}}'
%(x, \quad t) 
, \quad 
\bar E_+
%(x, \quad t) 
 = \hbar \bar\nu' 
%(x, \quad t) 
 + \bar e_{\mathrm{e}}
%(x, \quad t) 
, \quad 
\bar E_-
%(x, \quad t) 
 = \bar e_{\mathrm{e}}'
%(x, \quad t) 
, \quad 
E_-
%(x, \quad t) 
 = \hbar \nu '
%(x, \quad t) 
 + e_{\mathrm{e}}'
%(x, \quad t) 
, 
}%
\label{eq:45JQ} % \nonumber % \Z 
\end{eqnarray}}%
%+++++++++++++++++++++++++++++++++++++++++++++
%\end{widetext}%
where the arguments were dropped for brevity. Referring to these substitutions (or their suitable subsets) as ``c.v.'' we find, 
%=============================================
{\begin{eqnarray}\eqalign{ 
Z_C(i\mu_+,i\mu_+,-i\bar\mu_-,-i\bar\mu_-)\settoheight{\auxlv}{$|$}%
\raisebox{-0.3\auxlv}{$|_{{\mathrm{c.v.}}}$} 
%\\ 
= i\bar\mu G_{\mathrm{R}} d_{\mathrm{e}} - i \mu G_{\mathrm{R}}^* \bar d_{\mathrm{e}}
, 
}%
\label{eq:48JT} % \nonumber % \Z 
\end{eqnarray}}%
%+++++++++++++++++++++++++++++++++++++++++++++
cf.\ equation (\ref{eq:76NU}). Furthermore, 
%=============================================
{\begin{eqnarray}\eqalign{ \fl 
Z_C\ensuremath{\bigg(
\frac{\delta }{\delta E_+},
\frac{\delta }{\delta \bar E_+},
\frac{\delta }{\delta E_-},
\frac{\delta }{\delta \bar E_-}
 \bigg)}\settoheight{\auxlv}{$\Big|$}%
\raisebox{-0.3\auxlv}{$\Big|_{{\mathrm{c.v.}}}$} 
%\\ 
= -i 
\frac{\delta }{\delta e_{\mathrm{e}}'} 
G_{\mathrm{R}} 
\frac{\delta }{\delta \bar\nu' }
+i
\frac{\delta }{\delta \bar e_{\mathrm{e}}'} 
G_{\mathrm{R}}^*
\frac{\delta }{\delta \nu' }
, 
}%
\label{eq:46JR} % \nonumber % \Z 
\end{eqnarray}}%
%+++++++++++++++++++++++++++++++++++++++++++++
and, 
%=============================================
{\begin{eqnarray} 
\exp\ensuremath{\big(
i\bar\mu_+ E_+ 
+i\mu_+ \bar E_+ 
-i\bar\mu_- E_- 
-i\mu_- \bar E_-
 \big)}\settoheight{\auxlv}{$|$}%
\raisebox{-0.3\auxlv}{$|_{{\mathrm{c.v.}}}$} 
\nonumber\\ \qquad 
= \exp\ensuremath{\big(
i\bar\mu e_{\mathrm{e}}' 
+i\bar\nu' d_{\mathrm{e}} 
-i\mu\bar e_{\mathrm{e}}' 
-i\nu' \bar d_{\mathrm{e}} 
 \big)} . 
\label{eq:47JS} % \nonumber % \Z 
\end{eqnarray}}%
%+++++++++++++++++++++++++++++++++++++++++++++
Doubling of the phase space is best seen in equation (\ref{eq:48JT}) which expresses two independent fields, 
%=============================================
{\begin{eqnarray}\eqalign{ 
E(x,t) 
 = \int dx'dt' G_{\mathrm{R}}(x,x',t-t')d_{\mathrm{e}}(x',t') , \\ 
\bar E(x,t) 
 = \int dx'dt' G_{\mathrm{R}}^*(x,x',t-t')\bar d_{\mathrm{e}}(x',t') , 
}%
\label{eq:18WJ} % \nonumber % \Z 
\end{eqnarray}}%
%+++++++++++++++++++++++++++++++++++++++++++++
emitted by two independent sources, $d_{\mathrm{e}}(x,t)$ and $\bar d_{\mathrm{e}}(x,t)$. Such formal structure comes very handy if the goal of analyses is indeed derivation of generalised positive-P representations as in \cite{BWO}. This and forthcoming papers utilize a version of P-representation \cite{MandelWolf} rather than positive-P, so that doubling of auxiliary variables is better be avoided. 

To suppress doubling of the phase space we impose conditions on the causal variables, 
%=============================================
{\begin{eqnarray}\eqalign{ 
\bar\mu(x,t) = \mu ^*(x,t), \quad 
\bar d_{\mathrm{e}}(x,t) = d_{\mathrm{e}}^*(x,t), \\ 
\bar\nu'(x,t) = \nu^{\prime *}(x,t), \quad 
\bar e_{\mathrm{e}}'(x,t) = e_{\mathrm{e}}^{\prime *}(x,t). 
}%
\label{eq:78WC} % \nonumber % \Z 
\end{eqnarray}}%
%+++++++++++++++++++++++++++++++++++++++++++++
To preserve response substitutions (\ref{eq:45JQ}) we also have to impose conditions on the auxiliary variables, 
%=============================================
{\begin{eqnarray}\eqalign{ 
\bar\mu_{\pm}(x,t) = \mu_{\mp} ^*(x,t), \quad 
\bar E_{\pm}(x,t) = E_{\mp} ^*(x,t). 
}%
\label{eq:55CF} % \nonumber % \Z 
\end{eqnarray}}%
%+++++++++++++++++++++++++++++++++++++++++++++
Conditions (\ref{eq:78WC}), (\ref{eq:55CF}) are by definition consistent with response substitutions and do not interfere with above analyses, nor with analyses in paper \cite{WickCaus}. Wick's theorem (\ref{eq:38UK}) equally holds if independent functional derivatives by $E_{\pm}(x,t),\bar E_{\pm}(x,t)$ are replaced by pairs of complex-conjugate derivatives. All derivations in the above and in paper \cite{WickCaus} remain applicable. 

Noteworthy is that effective doubling of the phase space is also present in the broad-band\ case. Indeed, if we keep $\eta_{\pm}(x,t)$ real, $ j_{\mathrm{e}}(x,t)$ comes out complex, whereas a {\em classical source\/} $ j_{\mathrm{e}}(x,t)$ in (\ref{eq:15WE}) should be real. To keep the auxiliary variable $ j_{\mathrm{e}}(x,t)$ real, we could introduce the condition, 
%=============================================
{\begin{eqnarray}\eqalign{ 
\eta_-(x,t) = \eta_+^*(x,t) . 
}%
\label{eq:79WD} % \nonumber % \Z 
\end{eqnarray}}%
%+++++++++++++++++++++++++++++++++++++++++++++
However, this makes $\eta(x,t)$ purely imaginary, so that analytical extension to real $\eta(x,t)$ is needed. We leave the question of mathematical rigour behind response transformations open for discussion. 

%********************************************************
\subsection{Response transformation and solution in terms of dressed dipoles}%
\label{ch:RTX}
%********************************************************
Based on the above, we postulate response substitutions in the narrow-band\ case\ 
to be, (again dropping the arguments)
%\begin{widetext} 
%=============================================
{\begin{eqnarray}\eqalign{ 
\mu_+
%(x, \quad t) 
 = \frac{d_{\mathrm{e}}
%(x, \quad t) 
}{\hbar }, \quad 
\bar\mu_+
%(x, \quad t) 
 = \mu^*
%(x, \quad t) 
+ \frac{d_{\mathrm{e}}^*
%(x, \quad t) 
}{\hbar }, \quad 
\bar\mu_-
%(x, \quad t) 
 = \frac{d_{\mathrm{e}}^*
%(x, \quad t) 
}{\hbar }, \quad 
\mu_-
%(x, \quad t) 
 = \mu
%(x, \quad t) 
 + \frac{d_{\mathrm{e}}
%(x, \quad t) 
}{\hbar }, \quad 
\\ 
\nu_+
%(x, \quad t) 
 = \frac{e_{\mathrm{e}}
%(x, \quad t) 
}{\hbar }, \quad 
\bar\nu_+
%(x, \quad t) 
 = \nu^*
%(x, \quad t) 
+ \frac{e_{\mathrm{e}}^*
%(x, \quad t) 
}{\hbar }, \quad 
\bar\nu_-
%(x, \quad t) 
 = \frac{e_{\mathrm{e}}^*
%(x, \quad t) 
}{\hbar }, \quad 
\nu_-
%(x, \quad t) 
 = \nu
%(x, \quad t) 
 + \frac{e_{\mathrm{e}}
%(x, \quad t) 
}{\hbar }, 
\\ 
E_+
%(x, \quad t) 
 = e_{\mathrm{e}}'
%(x, \quad t) 
, \quad 
\bar E_+
%(x, \quad t) 
 = \hbar \nu^{\prime *} 
%(x, \quad t) 
+ e_{\mathrm{e}}^*
%(x, \quad t) 
, \quad 
\bar E_-
%(x, \quad t) 
 = e_{\mathrm{e}}^{\prime *}
%(x, \quad t) 
, \quad 
E_-
%(x, \quad t) 
 = \hbar \nu '
%(x, \quad t) 
 + e_{\mathrm{e}}'
%(x, \quad t) 
. 
}%
\label{eq:35YP} % \nonumber % \Z 
\end{eqnarray}}%
%+++++++++++++++++++++++++++++++++++++++++++++
The first and the third lines here are substitutions (\ref{eq:45JQ}) under conditions (\ref{eq:78WC}), (\ref{eq:55CF}). The second line is modelled on the first one; it imposes one more condition on auxiliary variables, 
%=============================================
{\begin{eqnarray}\eqalign{ 
\bar\nu_{\pm}(x,t) = \nu_{\mp} ^*(x,t). 
}%
\label{eq:54CE} % \nonumber % \Z 
\end{eqnarray}}%
%+++++++++++++++++++++++++++++++++++++++++++++
Similarly to the broad-band\ case\ we define the functionals, (cf.\ the closing remark in section \ref{ch:IC})
%=============================================
{\begin{eqnarray}\eqalign{ %\fl 
 \Phi\ensuremath{\big(
\mu ,\mu ^*,\nu ,\nu ^*
\big|
d_{\mathrm{e}},d_{\mathrm{e}}^*,e_{\mathrm{e}},e_{\mathrm{e}}^*
\big|
D_{\mathrm{e}},D_{\mathrm{e}}^*,E_{\mathrm{e}},E_{\mathrm{e}}^*
 \big)} 
\\ \qquad 
= \Xi\ensuremath{\big(
\bar\mu_+ ,\mu_+, 
\bar\mu_- ,\mu_-, 
\bar\nu_+ ,\nu_+, 
\bar\nu_- ,\nu_-
\big|E_{\mathrm{e}},E_{\mathrm{e}}^*,D_{\mathrm{e}},D_{\mathrm{e}}^* 
 \big)} 
\settoheight{\auxlv}{$|$}%
\raisebox{-0.3\auxlv}{$|_{\mathrm{c.v.}}$}, 
\\ %\fl 
 \Phi_{\mathrm{dev}}\ensuremath{\big(
\nu ,\nu ^*\big|e_{\mathrm{e}},e_{\mathrm{e}}^*
 \big)} 
%\\ 
 = \Phi\ensuremath{\big(
0,0,\nu ,\nu ^*
\big|
0,0,e_{\mathrm{e}},e_{\mathrm{e}}^*
\big|0,0,0,0
 \big)}, 
\\ %\fl 
 \Phi^{\mathrm{I}}_{\mathrm{dev}}\ensuremath{\big(
\nu ,\nu ^*\big|e_{\mathrm{e}},e_{\mathrm{e}}^*
 \big)} 
%\\ 
 = \Xi^{\mathrm{I}}_{\mathrm{dev}}\ensuremath{\big(
\bar\nu_+ ,\nu_+, 
\bar\nu_- ,\nu_-
 \big)} 
\settoheight{\auxlv}{$|$}%
\raisebox{-0.3\auxlv}{$|_{\mathrm{c.v.}}$} 
, 
}%
\label{eq:36YQ} % \nonumber % \Z 
\end{eqnarray}}%
%+++++++++++++++++++++++++++++++++++++++++++++
with ``c.v.'' referring to (suitable subsets of) equations (\ref{eq:35YP}). Rewriting (\ref{eq:37YR}) in causal variables we have, %=============================================
{\begin{eqnarray} \fl 
\Phi\ensuremath{\big(
\mu ,\mu ^*,\nu ,\nu ^*
\big|
d_{\mathrm{e}},d_{\mathrm{e}}^*,e_{\mathrm{e}},e_{\mathrm{e}}^*
\big|
D_{\mathrm{e}},D_{\mathrm{e}}^*,E_{\mathrm{e}},E_{\mathrm{e}}^*
 \big)} 
= \exp\ensuremath{\bigg[
-i
\frac{\delta }{\delta e_{\mathrm{e}}'} 
G_{\mathrm{R}} 
\frac{\delta }{\delta \nu^{\prime *} }
+ i 
\frac{\delta }{\delta e_{\mathrm{e}}^{\prime *}} 
G_{\mathrm{R}}^*
\frac{\delta }{\delta \nu' }
\bigg]}
\nonumber\\ \fl\qquad \times 
\exp\ensuremath{\big[
i\mu ^*e_{\mathrm{e}}'
-i\mu e_{\mathrm{e}}^{\prime *}
+i\nu ^{\prime *}\ensuremath{\big(
D_{\mathrm{e}}+d_{\mathrm{e}}
 \big)} 
-i\nu '\ensuremath{\big(
D_{\mathrm{e}}^*+d_{\mathrm{e}}^*
 \big)}
\big]} 
\nonumber\\ \fl\qquad\times 
\Phi_{\mathrm{dev}}^{\mathrm{I}}\ensuremath{\big(
\nu + \nu ',\nu ^*+\nu^{\prime *} 
\big|
E_{\mathrm{e}}+e_{\mathrm{e}}+e_{\mathrm{e}}'
, 
E_{\mathrm{e}}^*+e_{\mathrm{e}}^*+e_{\mathrm{e}}^{\prime *}
 \big)} 
\settoheight{\auxlv}{$|$}%
\raisebox{-0.3\auxlv}{$|_{\nu '=e_{\mathrm{e}}'=0}$}\, . 
\label{eq:38YS} % \nonumber % \Z 
\end{eqnarray}}%
%+++++++++++++++++++++++++++++++++++++++++++++
Confining this to dipoles we find the dressing relation, 
%=============================================
{\begin{eqnarray}\fl 
\Phi_{\mathrm{dev}}\ensuremath{\big(
\nu,\nu ^*, 
e_{\mathrm{e}},e_{\mathrm{e}}^*
 \big)} 
= \exp\ensuremath{\bigg[
-i
\frac{\delta }{\delta e_{\mathrm{e}}'} 
G_{\mathrm{R}} 
\frac{\delta }{\delta \nu^{\prime *} }
+ i 
\frac{\delta }{\delta e_{\mathrm{e}}^{\prime *}} 
G_{\mathrm{R}}^*
\frac{\delta }{\delta \nu' }
\bigg]}
%\nfp \times 
%\exp\Sbracket{\big}{
%i\mu ^*e_{\mathrm{e}}'
%-i\mu e_{\mathrm{e}}^{\prime *}
%+i\nu ^{\prime *}\Rbracket{\big}{
%D_{\mathrm{e}}+d_{\mathrm{e}}
%} 
%-i\nu '\Rbracket{\big}{
%D_{\mathrm{e}}^*+d_{\mathrm{e}}^*
%}
%}
\Phi_{\mathrm{dev}}^{\mathrm{I}}\ensuremath{\big(
\nu,\nu ^*, 
e_{\mathrm{e}},e_{\mathrm{e}}^*
 \big)} . 
\label{eq:40YU} % \nonumber % \Z 
\end{eqnarray}}%
%+++++++++++++++++++++++++++++++++++++++++++++
Using equation (\ref{eq:25YC}) and proceeding in close similarity to equations (\ref{eq:26YD})--(\ref{eq:28YF}) 
we find the solution in terms of dressed dipoles, 
%=============================================
{\begin{eqnarray} \fl 
\Phi\ensuremath{\big(
\mu ,\mu ^*,\nu ,\nu ^*
\big|
d_{\mathrm{e}},d_{\mathrm{e}}^*,e_{\mathrm{e}},e_{\mathrm{e}}^*
\big|
D_{\mathrm{e}},D_{\mathrm{e}}^*,E_{\mathrm{e}},E_{\mathrm{e}}^*
 \big)} %\\ 
= \exp\ensuremath{\big[
i\mu ^* G_{\mathrm{R}} \ensuremath{\big(
D_{\mathrm{e}}+d_{\mathrm{e}}
 \big)} 
- i\mu G_{\mathrm{R}}^* \ensuremath{\big(
D_{\mathrm{e}}^*+d_{\mathrm{e}}^*
 \big)} 
\big]} 
\nonumber\\ \fl\qquad \times 
\Phi_{\mathrm{dev}}\ensuremath{\big(
\nu+\mu G_{\mathrm{R}},\nu ^*+\mu^* G_{\mathrm{R}}^* 
\big | 
E_{\mathrm{e}}+e_{\mathrm{e}}+G_{\mathrm{R}}\ensuremath{\big(
d_{\mathrm{e}}+D_{\mathrm{e}}
 \big)} ,
E_{\mathrm{e}}^*+e_{\mathrm{e}}^*+G_{\mathrm{R}}^*\ensuremath{\big(
d_{\mathrm{e}}^*+D_{\mathrm{e}}^*
 \big)} 
 \big)} . \nonumber\\ \fl\qquad 
\label{eq:41YV} % \nonumber % \Z 
\end{eqnarray}}%
%+++++++++++++++++++++++++++++++++++++++++++++
%\end{widetext}%
The physical content of equations (\ref{eq:40YU}) and (\ref{eq:41YV}) is a subject of forthcoming papers. 
%********************************************************
\section{Putting the broad-band\ and the narrow-band\ case s together}%
\label{ch:UP}
%********************************************************
There is no difficulty whatsoever in unifying manipulations as in sections \ref{ch:I} and \ref{ch:EII} under the common umbrella of the model of section \ref{ch:GH} for arbitrary $M$ and $N$, $1<M<N$. These manipulations apply to different sets of modes, and do not interfere if united in a single problem. Formulae for the generic model of section \ref{ch:GH} may therefore be written straightway, by merging the corresponding formulae for the broad-band\ and narrow-band\ case s. So, the characteristic functional of the $T_C$-ordered products of the Heisenberg field, photocurrent and dipole operators, ${\hat{\mathcal A}}(x,t),{\hat{\mathcal E}}(x,t),{\hat{\mathcal J}}(x,t),{\hat{\mathcal D}}(x,t)$, is a merger of equations (\ref{eq:2XC}) and (\ref{eq:33YM}), (cf.\ the closing remark in section \ref{ch:IC})
%\begin{widetext} 
%=============================================
{\begin{eqnarray} %\fl 
\Phi\ensuremath{\big(
\eta ,
\mu ,\mu ^*, 
\zeta ,
\nu ,\nu ^*
\big | 
j_{\mathrm{e}},
d_{\mathrm{e}},d_{\mathrm{e}}^*, 
a_{\mathrm{e}},
e_{\mathrm{e}},e_{\mathrm{e}}^*
\big | 
J_{\mathrm{e}},
D_{\mathrm{e}},D_{\mathrm{e}}^*,
A_{\mathrm{e}},
E_{\mathrm{e}},E_{\mathrm{e}}^*
 \big)} 
\nonumber\\ \qquad 
= \ensuremath{\Big\langle T_C 
\exp\ensuremath{\big(
i{\eta}_+ {\hat{\mathcal A}}_+ 
-i{\eta}_- {\hat{\mathcal A}}_- 
+i{\zeta}_+ {\hat{\mathcal J}}_+ 
-i{\zeta}_- {\hat{\mathcal J}}_-
 \big)} 
\nonumber\\ \qquad\quad \times 
\exp\ensuremath{\big(
i\bar\mu_+ {\hat{\mathcal E}}_+ +i\mu_+ {\hat{\mathcal E}}_+^{\dag} 
-i\bar\mu_- {\hat{\mathcal E}}_- -i\mu_- {\hat{\mathcal E}}_-^{\dag}
\nonumber\\ \qquad\qquad 
+ i\bar\nu_+ {\hat{\mathcal D}}_+ +i\nu_+ {\hat{\mathcal D}}_+^{\dag} 
-i\bar\nu_- {\hat{\mathcal D}}_- -i\nu_- {\hat{\mathcal D}}_-^{\dag}
 \big)} 
\Big\rangle}\settoheight{\auxlv}{$\big|$}%
\raisebox{-0.3\auxlv}{$\big|_{\mathrm{c.v.}}$} , 
\label{eq:93FK} % \nonumber % \Z 
\end{eqnarray}}%
%+++++++++++++++++++++++++++++++++++++++++++++
where c.v.\ refers to merger of the broad-band\ and narrow-band\ response substitutions (\ref{eq:5VU}), (\ref{eq:50JV}) and (\ref{eq:35YP}). The properties of dressed photocurrents and dipoles are expressed by the functional, 
%=============================================
{\begin{eqnarray}\eqalign{ \fl
\Phi_{\mathrm{dev}}\ensuremath{\big(
\zeta ,\nu ,\nu ^*\big| a_{\mathrm{e}},e_{\mathrm{e}},e_{\mathrm{e}}^*
 \big)} %\\ 
= \Phi\ensuremath{\big(
0 ,0,0,\zeta, \nu ,\nu ^*
\big | 
0,
0,0, 
a_{\mathrm{e}},
e_{\mathrm{e}},e_{\mathrm{e}}^*
\big | 
0,0, 
0,0, 
0,0 
 \big)} . 
}%
\label{eq:90FF} % \nonumber % \Z 
\end{eqnarray}}%
%+++++++++++++++++++++++++++++++++++++++++++++
The formula reducing (\ref{eq:93FK}) to (\ref{eq:90FF})
is constructed merging equations (\ref{eq:28YF}) and (\ref{eq:41YV}), 
%=============================================
{\begin{eqnarray} \fl
\Phi\ensuremath{\big(
\eta ,
\mu ,\mu ^*, 
\zeta ,
\nu ,\nu ^*
\big | 
j_{\mathrm{e}},
d_{\mathrm{e}},d_{\mathrm{e}}^*, 
a_{\mathrm{e}},
e_{\mathrm{e}},e_{\mathrm{e}}^*
\big | 
J_{\mathrm{e}},
D_{\mathrm{e}},D_{\mathrm{e}}^*,
A_{\mathrm{e}},
E_{\mathrm{e}},E_{\mathrm{e}}^*
 \big)} 
\nonumber\\ \fl\qquad 
= \exp\ensuremath{\big[
i\eta\Delta_{\mathrm{R}} \ensuremath{\big( 
j_{\mathrm{e}} + J_{\mathrm{e}}
 \big)} 
+i\mu ^* G_{\mathrm{R}}\ensuremath{\big(
d_{\mathrm{e}}+D_{\mathrm{e}}
 \big)} 
-i\mu G_{\mathrm{R}}^*\ensuremath{\big(
d_{\mathrm{e}}^*+D_{\mathrm{e}}^*
 \big)}
\big]} 
\nonumber\\ \fl\qquad\quad\times 
\Phi_{\mathrm{dev}}\ensuremath{\big(
\zeta+\eta\Delta_{\mathrm{R}} ,\nu+\mu G_{\mathrm{R}}^* ,\nu^*+\mu^* G_{\mathrm{R}} 
\big | 
A_{\mathrm{ext\ tot}},E_{\mathrm{ext\ tot}},E_{\mathrm{ext\ tot}}^*
 \big)} , 
\label{eq:54XX} % \nonumber % \Z 
\end{eqnarray}}%
%+++++++++++++++++++++++++++++++++++++++++++++
where
%=============================================
{\begin{eqnarray}\eqalign{ %\fl
 A_{\mathrm{ext\ tot}}(x,t) = a_{\mathrm{e}}(x,t) 
+ A_{\mathrm{e}}(x,t) 
\\ \qquad + \int dx'dt' \Delta_{\mathrm{R}}(x,x',t-t') \ensuremath{\big[
j_{\mathrm{e}}(x',t')+J_{\mathrm{e}}(x',t')
\big]} , \\ %\fl
E_{\mathrm{ext\ tot}}(x,t) = e_{\mathrm{e}}(x,t) + E_{\mathrm{e}}(x,t) 
\\ \qquad + \int dx'dt' G_{\mathrm{R}}(x,x',t-t') \ensuremath{\big[
d_{\mathrm{e}}(x',t')+D_{\mathrm{e}}(x',t')
\big]} . 
}%
\label{eq:55XY} % \nonumber % \Z 
\end{eqnarray}}%
%+++++++++++++++++++++++++++++++++++++++++++++
%\end{widetext}%
Finally, the dressing relation is a merger of (\ref{eq:23YA}) and (\ref{eq:40YU})
%=============================================
{\begin{eqnarray} \fl 
\Phi_{\mathrm{dev}}\ensuremath{\big(
\zeta ,\nu,\nu ^* 
\big | 
a_{\mathrm{e}},e_{\mathrm{e}},e_{\mathrm{e}}^*
 \big)} %\nfp 
= 
\exp\ensuremath{\bigg[
-i 
\frac{\delta }{\delta a_{\mathrm{e}}} 
\Delta _{\mathrm{R}}
\frac{\delta }{\delta \zeta }
%\\ 
-i 
\frac{\delta }{\delta e_{\mathrm{e}}} 
G_{\mathrm{R}} 
\frac{\delta }{\delta \nu^* }
+i 
\frac{\delta }{\delta e_{\mathrm{e}}^*} 
G_{\mathrm{R}}^*
\frac{\delta }{\delta \nu }
\bigg]}
\nonumber\\ \qquad\qquad \times 
\Phi_{\mathrm{dev}}^{\mathrm{I}}\ensuremath{\big(
\zeta ,\nu,\nu ^* 
\big | 
a_{\mathrm{e}},e_{\mathrm{e}},e_{\mathrm{e}}^*
 \big)} , 
\label{eq:91FH} % \nonumber % \Z 
\end{eqnarray}}%
%+++++++++++++++++++++++++++++++++++++++++++++
with the formula for $\Phi_{\mathrm{dev}}^{\mathrm{I}}$ unifying (\ref{eq:85AT}), (\ref{eq:17XU}), (\ref{eq:90AY}), and (\ref{eq:36YQ}), 
%=============================================
{\begin{eqnarray} %\fl
\Phi_{\mathrm{dev}}^{\mathrm{I}}\ensuremath{\big(
\zeta ,\nu ,\nu ^* 
\big | 
a_{\mathrm{e}},e_{\mathrm{e}},e_{\mathrm{e}}^*
 \big)} 
%\nfp 
= \mathrm{Tr}\hat\rho_{\mathrm{dev}}
T_C\exp\ensuremath{\big(
i{\zeta}_+ \hat J_{+} 
-i{\zeta}_- \hat J_{-}
 \big)} 
\nonumber\\ \qquad\qquad \times 
\exp\ensuremath{\big(
i\bar\nu_+ \hat D_{+} +i\nu_+ \hat D_{+}^{\dag}
-i\bar\nu_- \hat D_{-} -i\nu_- \hat D_{-}^{\dag}
 \big)} 
\settoheight{\auxlv}{$\big|$}%
\raisebox{-0.3\auxlv}{$\big|_{\mathrm{c.v.}}$} . 
\label{eq:92FJ} % \nonumber % \Z 
\end{eqnarray}}%
%+++++++++++++++++++++++++++++++++++++++++++++
%********************************************************
\section{Conclusion and outlook}%
\label{ch:SO}
%********************************************************
General structure of electromagnetic interactions in response picture was elucidated. The astonishing feature of perturbative relations thus obtained is that they lack Planck's constant. It is hidden in functionals $\Phi^{\mathrm{I}}_{\mathrm{dev}}$ and $\Phi_{\mathrm{dev}}$ and should reappear were equations of motion for matter considered explicitly. However, {\em for electromagnetic interactions of ``dressed'' devices Planck's constant is irrelevant\/}. Implications of this point will be subject of forthcoming papers \cite{PFunc,Sudar}. 

\ack

Support of SFB/TRR 21 and of the Humboldt Foundation is gratefully acknowledged. 

\appendix
%********************************************************
\section{Diagram techniques}%
\label{ch:Diag}
%********************************************************
This appendix constitutes a formal closure of the paper, by associating dressing relations with a common computational method: Wyld-style diagram techniques \cite{Wyld}. It assumes a reader versed in functional techniques of quantum field theory \cite{VasF}. We consider the broad-band\ case. Our goal is to construct a generalised diagram expansion for the dressing relation (\ref{eq:23YA}). To this end, the functional $\Phi_{\mathrm{dev}}^{\mathrm{I}}$ is represented in terms of {\em bare response cumulants\/} $Q^{(m,n)}$, 
%=============================================
{\begin{eqnarray} \fl
\Phi_{\mathrm{dev}}^{\mathrm{I}}\ensuremath{\big(
\zeta 
\big | 
a_{\mathrm{e}}
 \big)} = \exp\ensuremath{\bigg[
\sum_{m,n=0}^{\infty}\frac{i^m}{m!n!}
\int dt_1\cdots dt_m 
dt_1'\cdots dt_n'
\nonumber\\ \fl\qquad \times 
Q^{(m,n)}\ensuremath{\big(
t_1,\cdots ,t_m \big | t_1',\cdots t_n'
 \big)} 
\zeta (t_1)\cdots\zeta (t_m)
 a_{\mathrm{e}}(t_1')\cdots a_{\mathrm{e}}(t_n')
\bigg]} . 
\label{eq:51JW} % \nonumber % \Z 
\end{eqnarray}}%
%+++++++++++++++++++++++++++++++++++++++++++++
For simplicity we drop all field arguments except time. The quantities $Q^{(m,n)}$ express properties of the bare current. For example, in spinor electrodynamics, they are found applying response transformations \cite{APII,APIII} to fermionic loops (implying that (\ref{eq:51JW}) is properly generalised to the 4-vector density of current). Similarly to $\Phi_{\mathrm{dev}}^{\mathrm{I}}$, functional $\Phi_{\mathrm{dev}}$ is represented in terms of {\em dressed response cumulants\/} $\mathcal{Q}^{(m,n)}$, 
%=============================================
{\begin{eqnarray} \fl
\Phi_{\mathrm{dev}}\ensuremath{\big(
\zeta 
\big | 
a_{\mathrm{e}}
 \big)} = \exp\ensuremath{\bigg[
\sum_{m,n=0}^{\infty}\frac{i^m}{m!n!}
\int dt_1\cdots dt_m 
dt_1'\cdots dt_n'
\nonumber\\ \fl\qquad \times 
\mathcal{Q}^{(m,n)}\ensuremath{\big(
t_1,\cdots ,t_m \big | t_1',\cdots t_n'
 \big)} 
\zeta (t_1)\cdots\zeta (t_m)
 a_{\mathrm{e}}(t_1')\cdots a_{\mathrm{e}}(t_n')
\bigg]} . 
\label{eq:53JY} % \nonumber % \Z 
\end{eqnarray}}%
%+++++++++++++++++++++++++++++++++++++++++++++
The quantities $\mathcal{Q}^{(m,n)}$ express properties of the dressed (observable, physical) current. Substituting (\ref{eq:51JW}) into (\ref{eq:23YA}), expanding all exponents in power series and comparing the result to (\ref{eq:53JY}) we recover the diagram series for the dressed cumulants. 
 
\newlength{\graphlength}
\settowidth{\graphlength}{\mbox{$\displaystyle M$}}
 
For the sake of argument, we assume that only three cumulants, $Q^{(1,1)}$, $Q^{(1,3)}$ and $Q^{(2,2)}$, are nonzero. In the diagram expansion, they serve as generalised vertices, 
%=============================================
{\begin{eqnarray}\eqalign{ 
Q^{(1,1)}(t|t') = \left\{
\raisebox{-0.45\graphlength}
{\includegraphics[scale=0.6]{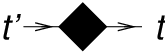}}\right\} , 
\\
Q^{(1,3)}(t|t_1',t_2',t_3') = \left\{
\raisebox{-1.63\graphlength}
{\includegraphics[scale=0.6]{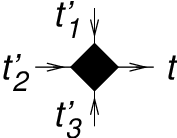}}\right\}, 
\\
Q^{(2,2)}(t_1,t_2|t_1',t_2') = \left\{
\raisebox{-1.85\graphlength}
{\includegraphics[scale=0.6]{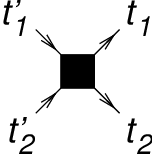}}\right\}. 
}%
\label{eq:52JX} % \nonumber % \Z 
\end{eqnarray}}%
%+++++++++++++++++++++++++++++++++++++++++++++
Vertices are distinguished by the number of incoming and outgoing lines; their orientation and other details are of no importance. Curly brackets isolate graphical elements visually. We also introduce a graphical notation for the causal propagator, 
%=============================================
{\begin{eqnarray}\eqalign{ 
\Delta_{\mathrm{R}}(t-t') = \left\{
\raisebox{-0.09\graphlength}
{\includegraphics[scale=0.6]{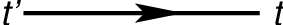}}\right\}. 
}%
\label{eq:56KB} % \nonumber % \Z 
\end{eqnarray}}%
%+++++++++++++++++++++++++++++++++++++++++++++
Then, for example, 
%=============================================
{\begin{eqnarray} \fl
\mathcal{Q}^{(1,1)}(t|t') 
= \left\{
\raisebox{-0.45\graphlength}
{\includegraphics[scale=0.6]{v11.eps}}\right\}
%\\ 
\nonumber\\ \fl\qquad 
+ \left\{
\raisebox{-0.45\graphlength}
{\includegraphics[scale=0.6]{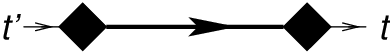}}\right\}
%\nnp 
+ \left\{
\raisebox{-1.55\graphlength}
{\includegraphics[scale=0.6]{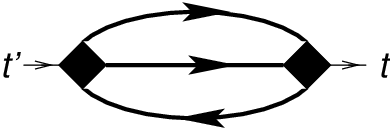}}\right\} +\cdots , 
\label{eq:54JZ} % \nonumber % \Z 
\end{eqnarray}}%
%+++++++++++++++++++++++++++++++++++++++++++++
%=============================================
{\begin{eqnarray}\eqalign{ \fl
\mathcal{Q}^{(1,3)}(t|t_1',t_2',t_3') = \left\{
\raisebox{-1.63\graphlength}
{\includegraphics[scale=0.6]{v13.eps}}\right\} 
%\\ 
+ \left\{
\raisebox{-1.8\graphlength}
{\includegraphics[scale=0.6]{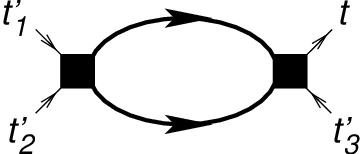}}\right\} +\cdots , 
}%
\label{eq:24WQ} % \nonumber % \Z 
\end{eqnarray}}%
%+++++++++++++++++++++++++++++++++++++++++++++
%=============================================
{\begin{eqnarray}\eqalign{ \fl
\mathcal{Q}^{(2,2)}(t_1,t_2|t_1',t_2') = \left\{
\raisebox{-1.85\graphlength}
{\includegraphics[scale=0.6]{v22.eps}}\right\} 
%\\ 
+ \left\{
\raisebox{-1.85\graphlength}
{\includegraphics[scale=0.6]{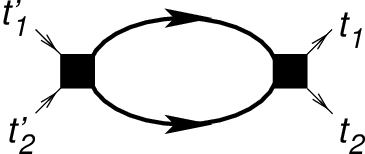}}\right\} +\cdots , 
}%
\label{eq:25WR} % \nonumber % \Z 
\end{eqnarray}}%
%+++++++++++++++++++++++++++++++++++++++++++++
etc., where 
%=============================================
{\begin{eqnarray}\eqalign{ \fl 
\left\{
\raisebox{-0.45\graphlength}
{\includegraphics[scale=0.6]{d11l.eps}}\right\} 
%\\ 
= \int d\bar t_1 d\bar t_2
Q^{(1,1)}(t|\bar t_1)\Delta_{\mathrm{R}}(\bar t_1 - \bar t_2)Q^{(1,1)}(\bar t_2|t'), 
}%
\label{eq:55KA} % \nonumber % \Z 
\end{eqnarray}}%
%+++++++++++++++++++++++++++++++++++++++++++++
%=============================================
{\begin{eqnarray} \fl
\left\{
\raisebox{-1.55\graphlength}
{\includegraphics[scale=0.6]{d11n.eps}}\right\} 
%\\ 
= \frac{1}{2}\int d\bar t_1 d\bar t_2d\bar t_3 d\bar t_4 d\bar t_5 d\bar t_6
\nonumber\\ \fl\qquad \times 
Q^{(2,2)}(t,\bar t_1|\bar t_2,\bar t_3)
%\\ \times 
\Delta_{\mathrm{R}}(\bar t_2 - \bar t_4)
\Delta_{\mathrm{R}}(\bar t_3 - \bar t_5)
\Delta_{\mathrm{R}}(\bar t_6 - \bar t_1)
Q^{(2,2)}(\bar t_4,\bar t_5|t',\bar t_6), 
\label{eq:21WM} % \nonumber % \Z 
\end{eqnarray}}%
%+++++++++++++++++++++++++++++++++++++++++++++
%=============================================
{\begin{eqnarray} \fl
\left\{
\raisebox{-1.85\graphlength}
{\includegraphics[scale=0.6]{d13.eps}}\right\}
= \frac{1}{2}\int d\bar t_1 d\bar t_2d\bar t_3 d\bar t_4
\nonumber\\ \fl\qquad \times 
Q^{(2,2)}(t|t'_3,\bar t_1,\bar t_2)
%\\ \times 
\Delta_{\mathrm{R}}(\bar t_1 - \bar t_3)
\Delta_{\mathrm{R}}(\bar t_2 - \bar t_4)
Q^{(2,2)}(\bar t_3,\bar t_4|t'_1,t'_2), 
\label{eq:23WP} % \nonumber % \Z 
\end{eqnarray}}%
%+++++++++++++++++++++++++++++++++++++++++++++
%=============================================
{\begin{eqnarray} \fl 
\left\{
\raisebox{-1.85\graphlength}
{\includegraphics[scale=0.6]{d22.eps}}\right\}
%\\ 
= \frac{1}{2}\int d\bar t_1 d\bar t_2d\bar t_3 d\bar t_4
\nonumber\\ \fl\qquad \times 
Q^{(2,2)}(t_1,t_2|\bar t_1,\bar t_2)
\Delta_{\mathrm{R}}(\bar t_1 - \bar t_3)
\Delta_{\mathrm{R}}(\bar t_2 - \bar t_4)
Q^{(2,2)}(\bar t_3,\bar t_4|t'_1,t'_2), 
\label{eq:22WN} % \nonumber % \Z 
\end{eqnarray}}%
%+++++++++++++++++++++++++++++++++++++++++++++
The general diagram rule is, match propagator outputs to vertex inputs, and {\em vice versa\/}, and integrate over all internal time arguments. 
By Mayer's first theorem \cite{VasF,Mayer}, series for response cumulants contain only connected diagrams. 
Coefficients at diagrams may be worked out following Vasil'ev \cite{VasF}. 
%********************************************************
\section{The narrow-band\ case\ as a resonance approximation to the broad-band\ case}%
\label{ch:UF}
%********************************************************
In this appendix we show that the narrow-band\ case\ may indeed be recovered by making the RWA in formulae for the broad-band\ case. This demonstrates consistency of the RWA with response transformations. Although as expected, this result is not automatic. For simplicity, we consider the case when the RWA is applied to all modes. In other words, we look at the change from $M=0$ to $M=N$. This will allow us to reuse notation of section \ref{ch:GH} with minimal amendments. Extension to the RWA applied only to a subset of modes is straightforward. 

For purposes of this section, the narrow-band\ field operators are given by equation (\ref{eq:4SX}) with $M\to N$, and the broad-band\ ones --- by equation (\ref{eq:45LB}) with $M\to 0$. Unlike in section \ref{ch:GH}, both operators are now defined with the same set of modes. The condition of resonance, 
%=============================================
{\begin{eqnarray}\eqalign{ 
|\omega _{\kappa }-\omega _0|\ll \omega _0, 
}%
\label{eq:13QJ} % \nonumber % \Z 
\end{eqnarray}}%
%+++++++++++++++++++++++++++++++++++++++++++++
now matters, so that $\hat E(x,t)$ and $\hat E^{\dag}(x,t)$ are indeed slow amplitudes. Using equation (\ref{eq:13QJ}) we can write simple relations between the nonresonant\ and resonant\ operators. In the resonance approximation, $\hat E (x,t)$ and $\hat E^{\dag} (x,t)$ are, up to factors, the frequency-positive and negative\ parts of $\hat A(x,t)$: 
%=============================================
{\begin{eqnarray}\eqalign{ 
\hat A(x,t) = \hat A^{(+)}(x,t)+\hat A^{(-)}(x,t), \\ 
\hat E (x,t)
 \approx 
i\omega_0\mathrm{e}^{i\omega _0t}\hat A^{(+)}(x,t) 
, \\ 
\hat E^{\dag} (x,t) 
 \approx 
-i\omega_0\mathrm{e}^{-i\omega _0t}\hat A^{(-)}(x,t) 
. 
}%
\label{eq:9QD} % \nonumber % \Z 
\end{eqnarray}}%
%+++++++++++++++++++++++++++++++++++++++++++++
By definition, we assume a similar structure for the current and dipole operators, 
%=============================================
{\begin{eqnarray}\eqalign{ 
 \hat J(x,t)= \hat J^{(+)} (x,t)+\hat J^{(-)}(x,t), \\ 
 {\eqalign{
\hat J^{(+)}(x,t) \approx -i\omega_0\mathrm{e}^{-i\omega _0t}\hat D(x,t), \\ 
\hat J^{(-)}(x,t) \approx i\omega_0\mathrm{e}^{i\omega _0t}\hat D^{\dag} (x,t). 
}} . 
}%
\label{eq:10QE} % \nonumber % \Z 
\end{eqnarray}}%
%+++++++++++++++++++++++++++++++++++++++++++++
$\hat D(x,t)$ and $\hat D^{\dag} (x,t)$ are also assumed to be varying slowly. Similar relations are postulated for the external sources; they are found from (\ref{eq:9QD}), (\ref{eq:10QE}) replacing $\hat A(x,t)\to A_{\mathrm{e}}(x,t)$, $\hat E(x,t)\to E_{\mathrm{e}}(x,t)$, $\hat J(x,t)\to J_{\mathrm{e}}(x,t)$, and $\hat D(x,t)\to D_{\mathrm{e}}(x,t)$. 

For the combination occuring in the $T$-exponents (\ref{eq:6XH}) and (\ref{eq:56ZM}) we have, 
%=============================================
{\begin{eqnarray}\eqalign{ \fl
\hat A\hat J+ A_{\mathrm{e}}\hat J+ \hat A J_{\mathrm{e}} 
%\\ 
= 
\hat E \hat D^{\dag} +E_{\mathrm{e}}\hat D ^{\dag}+\hat E\hat D_{\mathrm{e}}^* 
+\hat E^{\dag} \hat D+E_{\mathrm{e}}^*\hat D+\hat E^{\dag}\hat D_{\mathrm{e}} , 
}%
\label{eq:15QL} % \nonumber % \Z 
\end{eqnarray}}%
%+++++++++++++++++++++++++++++++++++++++++++++
where we use condensed notation (\ref{eq:3VS}). 
By itself, this relation is exact (subject to $\hat E(x,t)$ and $\hat D(x,t)$ being exactly frequency-positive, and $\hat E^{\dag}(x,t)$ and $\hat D^{\dag}(x,t)$ --- exactly frequency-negative). Cancellation of the counter-rotating terms is due to the formula, (with $f(t)$ and $g(t)$ being arbitrary functions)
%=============================================
{\begin{eqnarray}\eqalign{ 
\int_{-\infty}^{+\infty} dt f^{(+)}(t)g^{(+)}(t) = \int_{-\infty}^{+\infty} dt f^{(-)}(t)g^{(-)}(t) = 0,
}%
\label{eq:16QM} % \nonumber % \Z 
\end{eqnarray}}%
%+++++++++++++++++++++++++++++++++++++++++++++
which is in turn due to the standard formula for functions and their Fourier-images, 
%=============================================
{\begin{eqnarray}\eqalign{ 
\int_{-\infty}^{+\infty} dt f(t) g(t) = \int_{-\infty}^{+\infty} \frac{d\omega }{2\pi }f_{\omega }g_{-\omega } . 
}%
\label{eq:17QN} % \nonumber % \Z 
\end{eqnarray}}%
%+++++++++++++++++++++++++++++++++++++++++++++
In the $T$-exponents (\ref{eq:6XH}) and (\ref{eq:56ZM}), equation (\ref{eq:15QL}) survives only as an approximate relation, subject to the RWA. What prevents it from being exact under the time ordering are the theta-functions in equations (\ref{eq:10TD}) and (\ref{eq:11TE}). Clearly that (\ref{eq:15QL}) survives as an approximate relation under the orderings is just another form of the usual argument, that fast counter-rotating terms must average out in equations of motion for slow amplitudes. The narrow-band\ S-matrices (\ref{eq:30YJ}) and (\ref{eq:31YK}) thus indeed emerge as a resonance approximation to the broad-band\ ones (\ref{eq:6XH}) and (\ref{eq:56ZM}), as expected. 

Consider now the ``fate'' of the combination $\eta _+ \hat A_+ -\eta _- \hat A_-$ in equation (\ref{eq:5XF}) under response transformation (\ref{eq:5VU}). Disregarding the time orderings we have, 
%\begin{widetext} 
%=============================================
{\begin{eqnarray} \fl
\eta _+ \hat A_+ -\eta _- \hat A_- = \ensuremath{\bigg[
\frac{ j_{\mathrm{e}}}{\hbar }+\eta ^{(-)}
\bigg]} \ensuremath{\big[
\hat A_+^{(+)}+\hat A_+^{(-)}
\big]} - \ensuremath{\bigg[
\frac{ j_{\mathrm{e}}}{\hbar }-\eta ^{(+)}
\bigg]} \ensuremath{\big[
\hat A_-^{(+)}+\hat A_-^{(-)}
\big]} \nonumber\\ \fl\quad
= \frac{1}{i\omega _0}\int dx dt\ensuremath{\bigg\{
\mathrm{e}^{-i\omega _0t}\ensuremath{\bigg[
\frac{ j_{\mathrm{e}}^{(-)}(x,t)}{\hbar } + \eta ^{(-)}(x,t)
\bigg]} \hat E_+(x,t)
-\frac{\mathrm{e}^{i\omega _0t} j_{\mathrm{e}}^{(+)}(x,t)}{\hbar }\hat E^{\dag}_+(x,t)
\nonumber\\ \fl\qquad\quad 
-\frac{\mathrm{e}^{-i\omega _0t} j_{\mathrm{e}}^{(-)}(x,t)}{\hbar }\hat E_-(x,t)
+\mathrm{e}^{i\omega _0t}\ensuremath{\bigg[
\frac{ j_{\mathrm{e}}^{(+)}(x,t)}{\hbar } - \eta ^{(+)}(x,t)
\bigg]} \hat E_-^{\dag}(x,t)
\bigg\}} . 
\label{eq:18QP} % \nonumber % \Z 
\end{eqnarray}}%
%+++++++++++++++++++++++++++++++++++++++++++++
Because of the exponential factors condensed notation (\ref{eq:3VS}) becomes inapplicable, forcing us to write the last expression in full notation. 
Again, by itself equation (\ref{eq:18QP}) is exact; it turns approximate under the orderings. Comparing equation (\ref{eq:18QP}) to (\ref{eq:34YN}) leads us into defining, 
%=============================================
{\begin{eqnarray}\eqalign{ 
\fl\mu _+(x,t) = -\frac{\mathrm{e}^{i\omega _0t}}{i\omega _0}
\frac{j_{\mathrm{e}}^{(+)}(x,t)}{\hbar }, 
\quad 
\bar\mu _+(x,t) = \frac{\mathrm{e}^{-i\omega _0t}}{i\omega _0}\ensuremath{\bigg[
\eta ^{(-)}(x,t) +\frac{ j_{\mathrm{e}}^{(-)}(x,t)}{\hbar } 
\bigg]} , 
\\ \fl 
\bar\mu _-(x,t) = 
\frac{\mathrm{e}^{-i\omega _0t}}{i\omega _0} 
\frac{j_{\mathrm{e}}^{(-)}(x,t)}{\hbar } 
, 
\quad 
\mu _-(x,t) = \frac{\mathrm{e}^{i\omega _0t}}{i\omega _0}\ensuremath{\bigg[
\eta ^{(+)}(x,t) - \frac{ j_{\mathrm{e}}^{(+)}(x,t)}{\hbar }
\bigg]} . 
}%
\label{eq:19QQ} % \nonumber % \Z 
\end{eqnarray}}%
%+++++++++++++++++++++++++++++++++++++++++++++
%\end{widetext}%
In turn, comparing this to the first line of (\ref{eq:35YP}) we find the RWA correspondences for the auxiliary variables, 
%=============================================
{\begin{eqnarray}\eqalign{ 
d_{\mathrm{e}}(x,t) = -\frac{\mathrm{e}^{i\omega _0t} j_{\mathrm{e}}^{(+)}(x,t)}{i\omega _0}, \quad 
d_{\mathrm{e}}^*(x,t) = \frac{\mathrm{e}^{-i\omega _0t} j_{\mathrm{e}}^{(-)}(x,t)}{i\omega _0} , 
\\ 
\mu (x,t) = \frac{\mathrm{e}^{i\omega _0t} \eta ^{(+)}(x,t)}{i\omega _0}, 
\mu ^*(x,t) = \frac{\mathrm{e}^{-i\omega _0t} \eta ^{(-)}(x,t)}{i\omega _0} . 
}%
\label{eq:20QR} % \nonumber % \Z 
\end{eqnarray}}%
%+++++++++++++++++++++++++++++++++++++++++++++
With these definitions, 
%=============================================
{\begin{eqnarray} 
\eta _+ \hat A_+ - \eta _- \hat A_-
=
\bar\mu _+ \hat E_+ - \mu _+ \hat E_+^{\dag}
+ \bar\mu _- \hat E_- - \mu _- \hat E_-^{\dag} \nonumber\\ \qquad 
= 
\mu ^* \hat E_+ - \mu \hat E_-^{\dag} + \frac{d_{\mathrm{e}}^*\ensuremath{\big(
\hat E_++\hat E_-
 \big)} }{\hbar } + \frac{d_{\mathrm{e}}\ensuremath{\big(
\hat E_+^{\dag} +\hat E_-^{\dag}
 \big)} }{\hbar } 
. 
\label{eq:23QU} % \nonumber % \Z 
\end{eqnarray}}%
%+++++++++++++++++++++++++++++++++++++++++++++
The last formula here prepares ground for the response transformation of Wick's theorem. By similar arguments, 
%=============================================
{\begin{eqnarray}\eqalign{ 
e_{\mathrm{e}}(x,t) = \frac{\mathrm{e}^{i\omega _0t} a_{\mathrm{e}}^{(+)}(x,t)}{i\omega _0}, \quad 
e_{\mathrm{e}}^*(x,t) = -\frac{\mathrm{e}^{-i\omega _0t} a_{\mathrm{e}}^{(-)}(x,t)}{i\omega _0} , 
\\ 
\nu (x,t) = \frac{\mathrm{e}^{i\omega _0t} \zeta ^{(+)}(x,t)}{i\omega _0}, \quad 
\nu ^*(x,t) = \frac{\mathrm{e}^{-i\omega _0t} \zeta ^{(-)}(x,t)}{i\omega _0} , 
}%
\label{eq:24QV} % \nonumber % \Z 
\end{eqnarray}}%
%+++++++++++++++++++++++++++++++++++++++++++++
and 
%=============================================
{\begin{eqnarray} 
\eta _+ \hat J_+ - \eta _- \hat J_-
=
\bar\nu _+ \hat D_+ - \nu _+ \hat D_+^{\dag}
+ \bar\nu _- \hat D_- - \nu _- \hat D_-^{\dag} \nonumber\\ \qquad 
= 
\nu ^* \hat D_+ - \nu \hat D_-^{\dag} + \frac{e_{\mathrm{e}}^*\ensuremath{\big(
\hat D_++\hat D_-
 \big)} }{\hbar } + \frac{e_{\mathrm{e}}\ensuremath{\big(
\hat D_+^{\dag} +\hat D_-^{\dag}
 \big)} }{\hbar } 
, 
\label{eq:25QW} % \nonumber % \Z 
\end{eqnarray}}%
%+++++++++++++++++++++++++++++++++++++++++++++
where the relation between $\bar\mu _{\pm}(x,t)$ and $\nu (x,t),e_{\mathrm{e}}(x,t)$ is by definition given by the second line in equation (\ref{eq:35YP}). 

These formulae exhibit a number of important consistencies. The connection between $d_{\mathrm{e}}(x,t)$ and $j_{\mathrm{e}}(x,t)$ coincides with that between $\hat D(x,t)$ and $\hat J(x,t)$ and {\em ipso facto\/} with that between $D_{\mathrm{e}}(x,t)$ and $J_{\mathrm{e}}(x,t)$. The property that $J_{\mathrm{e}}(x,t)$ and $j_{\mathrm{e}}(x,t)$ occur as a sum thus neatly turns into the property that $D_{\mathrm{e}}(x,t)$ and $d_{\mathrm{e}}(x,t)$ occur as a sum. Similar consitencies exist among pairs $\hat E(x,t),\hat A(x,t)$, $E_{\mathrm{e}}(x,t),A_{\mathrm{e}}(x,t)$ and $e_{\mathrm{e}}(x,t),a_{\mathrm{e}}(x,t)$. Furthermore, equations (\ref{eq:20QR}) force $j_{\mathrm{e}}(x,t)$ to be real, and $\eta (x,t)$ to be purely imaginary. This agrees with the remarks in the last paragraph of section \ref{ch:RC}. The same applies to $a_{\mathrm{e}}(x,t)$ and $\nu (x,t)$ in equations (\ref{eq:24QV}). 

To conclude our argument we look at the reordering forms. Using equations (\ref{eq:20FA}) and \ref{eq:88LM}) it is easy to see that, 
%=============================================
{\begin{eqnarray}\eqalign{ \fl
\Delta_{\mathrm{R}}(x,x',t-t') \approx \frac{1}{\omega _0^2}\ensuremath{\big[
\mathrm{e}^{-i\omega _0(t-t')}G_{\mathrm{R}}(x,x',t-t') 
+ 
\mathrm{e}^{i\omega _0(t-t')}G_{\mathrm{R}}^*(x,x',t-t')
\big]} . 
}%
\label{eq:21QS} % \nonumber % \Z 
\end{eqnarray}}%
%+++++++++++++++++++++++++++++++++++++++++++++
Making use of equations (\ref{eq:20QR}) and dropping nonresonant\ terms we find, 
%=============================================
{\begin{eqnarray}\eqalign{ 
\eta \Delta_{\mathrm{R}} j_{\mathrm{e}} \approx 
\mu ^* G_{\mathrm{R}} d_{\mathrm{e}}
- \mu G_{\mathrm{R}}^* d_{\mathrm{e}}^* . 
}%
\label{eq:22QT} % \nonumber % \Z 
\end{eqnarray}}%
%+++++++++++++++++++++++++++++++++++++++++++++
Unlike (\ref{eq:15QL}) and (\ref{eq:18QP}), this relation is approximate because of the theta-function in $\Delta_{\mathrm{R}}$ and $G_{\mathrm{R}}$. 

Extending equation (\ref{eq:22QT}) to quadratic forms of derivatives in equation (\ref{eq:19XW}) and (\ref{eq:23YA}) takes a bit of mathematical effort. In the genuine narrow-band\ case\ of section \ref{ch:EII}, functional variables $\nu (x,t)$ and $e_{\mathrm{e}} (x,t)$ are unconstrained. Whereas their namesakes introduced by equation (\ref{eq:24QV}) are, up to a factor, the frequency-positive\ parts of other variables. To show, for instance, that equation (\ref{eq:23YA}) turns under the RWA into (\ref{eq:40YU}), we need a bridging concept of constrained derivative by the frequency-positive and negative\ parts of a function. We define it by the formula, with $f(x,t)$ being an arbitrary functional variable, 
%=============================================
{\begin{eqnarray}\eqalign{ 
\frac{\delta }{\delta f ^{(\pm)}(x,t)} 
= \ensuremath{\bigg[
\frac{\delta }{\delta f(x,t)}
\bigg]}^{(\mp)} 
%\\ 
%= \int dt' \delta \FPM(x,t'-t)
%\frac{\delta }{\delta f(x,t')} 
, 
}%
\label{eq:29KK} % \nonumber % \Z 
\end{eqnarray}}%
%+++++++++++++++++++++++++++++++++++++++++++++
so that, 
%=============================================
{\begin{eqnarray}\eqalign{ 
\frac{\delta }{\delta f(x,t)} 
= \frac{\delta }{\delta f ^{(+)}(x,t)} + 
\frac{\delta }{\delta f ^{(-)}(x,t)} . 
}%
\label{eq:84SB} % \nonumber % \Z 
\end{eqnarray}}%
%+++++++++++++++++++++++++++++++++++++++++++++
This definition follows the observation that separation of the frequency-positive and negative\ parts of a function, 
 
%=============================================
{\begin{eqnarray}\eqalign{ 
 f(x,t) = f^{(+)}(x,t) + f^{(-)}(x,t) , 
}%
\label{eq:82RZ} % \nonumber % \Z 
\end{eqnarray}}%
%+++++++++++++++++++++++++++++++++++++++++++++
is in fact an orthogonal decomposition. 
Indeed, for any pair of functions, 
%=============================================
{\begin{eqnarray}\eqalign{ 
\int dt f ^{(+)}(x,t)\ensuremath{\big[
g ^{(-)}(x',t)
\big]} ^* = 0 . 
}%
\label{eq:83SA} % \nonumber % \Z 
\end{eqnarray}}%
%+++++++++++++++++++++++++++++++++++++++++++++
This formula may be verified using equation (\ref{eq:17QN})). Furthermore, the $^{(\pm)}$ operations are projections. The frequency-positive\ part of a frequency-positive\ function is this function, the frequency-positive\ part of a frequency-negative\ function is zero, and so on. Definition (\ref{eq:29KK}) naturally extends the said orthogonal decomposition to the derivative $\delta /\delta f(x,t)$. In particular, for any functional $F(\cdot)$, 
%=============================================
{\begin{eqnarray} \fl 
\delta F(f) = 
\int dx dt \frac{\delta F(f)}{\delta f(x,t)}\delta f(x,t) 
\nonumber\\ \fl\qquad
= 
\int dx dt \ensuremath{\bigg[
\frac{\delta F(f)}{\delta f^{(+)}(x,t)}\delta f^{(+)}(x,t) 
+\frac{\delta F(f)}{\delta f^{(-)}(x,t)}\delta f^{(-)}(x,t)
\bigg]} . 
\label{eq:27QY} % \nonumber % \Z 
\end{eqnarray}}%
%+++++++++++++++++++++++++++++++++++++++++++++
Algebraic manipulation of derivatives (\ref{eq:29KK}) reduces to using their definition and observing that $\delta /\delta f^{(+)}(x,t)$ and $\delta /\delta f^{(-)}(x,t)$ are frequency-negative\ and frequency-positive, respectively. It is then straightforward to show that 
%=============================================
{\begin{eqnarray}\eqalign{ 
\frac{\delta }{\delta a_{\mathrm{e}}}\Delta_{\mathrm{R}} 
\frac{\delta }{\delta \zeta }
\approx 
\frac{\delta }{\delta e_{\mathrm{e}}}G_{\mathrm{R}}
\frac{\delta }{\delta \nu^* }
- 
\frac{\delta }{\delta e_{\mathrm{e}}^*}G_{\mathrm{R}}^* 
\frac{\delta }{\delta \nu }
, 
}%
\label{eq:26QX} % \nonumber % \Z 
\end{eqnarray}}%
%+++++++++++++++++++++++++++++++++++++++++++++
so that equation (\ref{eq:40YU}) is indeed a resonance approximation to (\ref{eq:23YA}). All formulae for the narrow-band\ case\ may thus be obtained making the resonance approximation in the formulae for the broad-band\ case. 

Note that, strictly speaking, we have demonstrated that the RHS's of all relations in section \ref{ch:EII} are the RWA to the RHS's of the corresponding relations in section \ref{ch:I}. This is obviously consistent with the assumptions one normally makes about their LHS's. For instance, it is assumed that equations (\ref{eq:9QD}) for the interaction-picture operators extend to the Heisenberg operators, 
%=============================================
{\begin{eqnarray}\eqalign{ 
{\hat{\mathcal A}}(x,t) = {\hat{\mathcal A}}^{(+)}(x,t)+{\hat{\mathcal A}}^{(-)}(x,t), \\ 
{\hat{\mathcal E}} (x,t)
 \approx 
i\omega_0\mathrm{e}^{i\omega _0t}{\hat{\mathcal A}}^{(+)}(x,t) 
, \\ 
{\hat{\mathcal E}}^{\dag} (x,t) 
 \approx 
-i\omega_0\mathrm{e}^{-i\omega _0t}{\hat{\mathcal A}}^{(-)}(x,t) 
. 
}%
\label{eq:28QZ} % \nonumber % \Z 
\end{eqnarray}}%
%+++++++++++++++++++++++++++++++++++++++++++++
The accuracy with which this formula holds is exactly the accuracy with which equation (\ref{eq:15QL}) holds under the time orderings. The same applies to equation (\ref{eq:10QE}) generalised to the Heisenberg operators.

%\newcommand{\Macros}{../../../Macros/} 
%%*******************************************
%\bibliography{\Macros strbin,QDynResp,\Macros notebin,\Macros citebin}

\begin{thebibliography}{10}
\bibitem{endTrans}
Statements like this one should be taken with a grain of salt. Transients due
  to switching the interaction on and off are infinite for any realistic model
  of the field. Yet more infinities are brought about by point-like atoms.
  Strictly speaking, Fermi's arrangement does not afford a consistent
  description in quantum electrodynamics, except in the lowest order of
  perturbation theory
\bibitem{endMeso}
The term ``mesoscopic'' refers to quantum manifestations under macroscopic
  conditions. Common as it actually is, this term is confusing. It implies that
  there exists something ``else'' that distinguishes macroscopic and mesoscopic
  conditions. In fact, there is nothing ``else'' apart from the hard-working
  experimental physicist, rescuing fragile quantum effects from being destroyed
  by environmental and measurement noises. We regard these terms as synonyms,
  with a tendency to apply ``macroscopic'' to experiment and
  ``mesoscopic''---to theory (as in ``macroscopic experiments are described by
  mesoscopic phenomenologies'')
\end{thebibliography}
%*******************************************
\providecommand{\newblock}{}

%******************************************* 
\end{document}